\documentclass[12pt]{article}
\usepackage{a4}
\usepackage{array}

\chardef\tempcat=\the\catcode`\@
\catcode`\@=11

\def\@gobble#1{}
\def\@testgrave{\`}
\def\@stressit{\futurelet\chartest\@stresschar }

\def\@stresschar#1{%
  \ifx #1y\def\result{\futurelet\chartest\@yligature}%
  \else \ifx #1Y\def\result{\futurelet\chartest\@Yligature}%
  \else \ifx\chartest\@testgrave \def\result{\accent"26 }%
  \else \def\result{\accent"26 #1}%
  \fi \fi \fi
  \result }

\def\@yligature{%
  \ifx a\chartest \def\result{\accent"26 \char"1F \@gobble}%
  \else \ifx u\chartest \def\result{\accent"26 \char"18 \@gobble}%
  \else \def\result{\accent"26 y}%
  \fi \fi
  \result }

\def\@Yligature{%
  \ifx a\chartest \def\result{\accent"26 \char"17 \@gobble}%
  \else \ifx A\chartest \def\result{\accent"26 \char"17 \@gobble}%
  \else \ifx u\chartest \def\result{\accent"26 \char"10 \@gobble}%
  \else \ifx U\chartest \def\result{\accent"26 \char"10 \@gobble}%
  \else \def\result{\accent"26 Y}%
  \fi \fi \fi \fi
  \result }

\def\!{\ifmmode \mskip-\thinmuskip \fi}

\def\cyracc{\chardef\i="10%
  \def\cydot{{\kern0pt}}%
  \def\cprime{\char"7E }\def\Cprime{\char"5E }%
  \def\cdprime{\char"7F }\def\Cdprime{\char"5F }%
  \def\dbar{dj}\def\Dbar{Dj}%
  \def\dz{\char"1E }\def\Dz{\char"16 }%
  \def\dzh{\char"0A }\def\Dzh{\char"02 }%
  \def\'##1{\if c##1\char"0F %
    \else \if C##1\char"07 %
    \else \accent"26 ##1\fi \fi }%
  \def\`##1{\if e##1\char"0B %
    \else \if E##1\char"03 %
    \else \errmessage{accent \string\` not defined in cyrillic}%
        ##1\fi \fi }%
  \def\=##1{\if e##1\char"0D %
    \else \if E##1\char"05 %
    \else \if \i##1\char"0C %
    \else \if I##1\char"04 %
    \else \errmessage{accent \string\= not defined in cyrillic}%
        ##1\fi \fi \fi \fi }%
  \def\u##1{\if \i##1\accent"24 i%
    \else \accent"24 ##1\fi }%
  \def\"##1{\if \i##1\accent"20 \char"3D %
    \else \if I##1\accent"20 \char"04 %
    \else \accent"20 ##1\fi \fi }%
  \def\!{\ifmmode \def\result{\mskip-\thinmuskip}%
    \else \def\result{\@stressit}\fi \result}}

\def\keep@cyracc{\let\cyr=\relax \let\i=\relax
        \let\ubar=\relax \let\cydot=\relax
        \let\cprime=\relax \let\Cprime=\relax
        \let\cdprime=\relax \let\Cdprime=\relax
        \let\dbar=\relax \let\Dbar=\relax
        \let\dz=\relax \let\Dz=\relax
        \let\dzh=\relax \let\Dzh=\relax
        \let\'=\relax \let\`=\relax \let\==\relax
        \let\u=\relax \let\"=\relax \let\!=\relax }

\DeclareFontEncoding{OT2}{\cyracc}{}

\catcode`\@=\tempcat

\newlength{\adressabstand}
\input{diagrams}
\newarrow{auf}|---{->}
\newarrow{nach}----{->}
\newarrow{ident}33333
\newarrow{aufsurj}|---{->>}
\newarrow{nachsurj}----{->>}
\newlength{\CDhoehe}                  
\newlength{\CDgap}                    
\usepackage{ifthen,array}
\newcommand{\ams}{\usepackage{amsfonts,amssymb,amsmath}}

\ams
\allowdisplaybreaks[3]
\newlength{\textwidthorig}
\newlength{\oddsidemarginorig}
\newlength{\textheightorig}
\newlength{\topmarginorig}
\def\seitenlaengenabsolut#1 #2 #3 #4 {\setlength{\textwidth}{#1}
                                      \setlength{\oddsidemargin}{#2}
                                      \setlength{\textheight}{#3}
                                      \setlength{\topmargin}{#4}}
\def\seitenlaengenrelzustandard#1 #2 #3 #4 {\setlength{\textwidth}{\textwidthorig+#1}
                                            \setlength{\oddsidemargin}{\oddsidemarginorig+#2}
                                            \setlength{\textheight}{\textheightorig+#3}
                                            \setlength{\topmargin}{\topmarginorig+#4}}
\def\seitenlaengenrelzuvorher#1 #2 #3 #4 {\addtolength{\textwidth}{#1}
                                          \addtolength{\oddsidemargin}{#2}
                                          \addtolength{\textheight}{#3}
                                          \addtolength{\topmargin}{#4}}
\newcommand{\standardseite}{\seitenlaengenrelzuvorher2.2cm -0.8cm 1.8cm -1.5cm }   
\standardseite
\newcommand{\Wegdamit}[1]{}
\newcommand{\leerezeile}{\vspace{2ex}}
\newcommand{\nach}{\longrightarrow}      
\newcommand{\txtnach}[1]{\xrightarrow{#1}}
\newcommand{\auf}{\longmapsto}           
\newcommand{\txtauf}[1]{\auf}            
\newcommand{\impliz}{\Longrightarrow}    
\newcommand{\aequ}{\Longleftrightarrow}  
 
\newcommand{\invimpliz}{\Longleftarrow}  
\newcommand{\iso}{\cong}                 
\newcommand{\ident}{\equiv}              
\newcommand{\teilmenge}{\subseteq}       
\newcommand{\obermenge}{\supseteq}       
\newcommand{\echteobermenge}{\supset}    
\newcommand{\aeqrel}{\sim}               

\newcommand{\nichtleq}{\text{$\leq\hspace{-2.3ex}/\hspace{1.0ex}$}}
\newcommand{\fueralle}{\hspace{1.7em}\forall}
\newcommand{\leeremenge}{\emptyset}      
\newcommand{\kreuz}{\times}              

\newcommand{\einschr}[1]{\mid_{#1}}      

\newcommand{\clos}[1]{\text{Cl}_{#1}}    
\newcommand{\betraganpass}[1]%
           {\left| #1 \right|}           
\newcommand{\betrag}[1]%
           {\betraganpass{#1}}           
\newcommand{\betragnichtanpass}[1]%
           {\mid #1 \mid}                
\newcommand{\norm}[1]%
           {\parallel #1 \parallel}      
\newcommand{\erww}[1]%
           {\langle #1 \rangle}          
\newcommand{\skalprod}[2]%
           {\langle #1,#2 \rangle}       
\newcommand{\quer}{\overline}            
\newcommand{\inv}[1]{\frac{1}{#1}}       
\newcommand{\im}{\text{im\;}}                          
\newcommand{\id}{\:\text{id}}                          
\newcommand{\inter}{\text{int}\:}                      
\newcommand{\diam}{\text{diam }}                       
\newcommand{\dist}{\text{dist}}                        
\newcommand{\Ad}{{\text{Ad}}}                          
\newcommand{\elanz}{\#}                                
\newcommand{\del}{\partial}                            
\newcommand{\Hom}{\text{Hom}}                          
\newcommand{\field}[1]{\mathbb{#1}}                    
\newcommand{\N}{{\field{N}}}                           
\newcommand{\R}{{\field{R}}}                           
\newcommand{\rnkl}[2]{\raisebox{-0.5ex}{$#1$}%
\raisebox{-0.2ex}{{\Large$\setminus$}}\,#2}            
\newcommand{\agb}{{\overline{{\cal A}/{\cal G}}}}      
\newcommand{\agbfact}[1][]{\text{$\agb/\!\aeqrel$}}    
\newcommand{\ag}{{\cal A}/{\cal G}}                    
\newcommand{\Ab}{{\overline{{\cal A}}}}                
\newcommand{\A}{{\cal A}}                              
\newcommand{\Gb}{{\overline{{\cal G}}}}                
\newcommand{\G}{{\cal G}}                              
\newcommand{\AbGb}{{\Ab/\Gb}}                          
\newcommand{\qa}{{\quer{A}}}                           
\newcommand{\qg}{{\quer{g}}}                           
\newcommand{\hg}{{\cal HG}}                            

\newcommand{\holgr}{{\mathbf H}}                       
\newcommand{\bz}{{\mathbf B}}                          
\newcommand{\eo}{{\mathbf E}}                          
\newcommand{\GR}{\Gamma}                               
\newcommand{\Ver}{\mathbf{V}}                          
\newcommand{\Edg}{\mathbf{E}}                          
\newcommand{\gross}[1]{{\boldsymbol #1}}               
\newcommand{\ga}{\gross{\alpha}}                       
\newcommand{\pf}[1]{{\cal P}_{#1}}                     
\newcommand{\Pf}{{\cal P}}                             
\newcommand{\BB}{\uparrow\uparrow}                     
\newcommand{\EB}{\downarrow\uparrow}                   
\newcommand{\notBB}{\mbox{${}\BB{}$\hspace*{-2.8ex}\rule[0.40ex]%
                    {2ex}{0.4pt}\hspace*{0.8ex}}}      
\newcommand{\notEB}{\mbox{${}\EB{}$\hspace*{-2.8ex}\rule[0.40ex]%
                    {2ex}{0.4pt}\hspace*{0.8ex}}}      
\newcommand{\Haar}{{\text{Haar}}}                      
\newcommand{\LG}{{\mathbf{G}}}                         
\newcommand{\aeqrelzush}[1][]{\sim}                    
\newcommand{\Abgeq}[1]{\Ab_{\geq#1}}                   
\newcommand{\Abeq}[1]{\Ab_{=#1}}                       
\newcommand{\Ableq}[1]{\Ab_{\leq#1}}                   
\newcommand{\typ}{\text{Typ}}                          
\newcommand{\bweg}{e}                                  
\newcommand{\stratallg}{{\cal S}}                      
\newcommand{\strat}{{\cal S}}                          
\newcommand{\qS}{{\quer S}}                            
\newcommand{\qz}{{\quer z}}                            
\newcommand{\T}{{\cal T}}
\newcommand{\nklza}[1][]{\ifthenelse{\equal{#1}{}}     
                                    {\rnkl{Z(\holgr_\qa)}{\LG}}        
                                   {\rnkl{Z(\holgr_{#1})}{\LG}}}       
\newcommand{\nkla}[1][]{\ifthenelse{\equal{#1}{}}      
                                    {\rnkl{\bz(\qa)}{\Gb}}        
                                    {\rnkl{\bz(#1)}{\Gb}}}       
\newcommand{\vect}[1]{\vec{#1}}      
\newcommand{\ListNullAbstaende}{\setlength{\topsep}{0pt}%
                                \setlength{\parskip}{0pt}%
                                \setlength{\partopsep}{0pt}%
                                \setlength{\itemsep}{0pt}%
                                \setlength{\parsep}{0pt}}
\newcommand{\ListNurAnstrichAbstand}{\setlength{\topsep}{0pt}%
                                     \setlength{\parskip}{0pt}%
                                     \setlength{\partopsep}{0pt}%
                                     \setlength{\parsep}{0pt}}
\newenvironment{StandardListe}[2]%
               {\begin{list}%
                      {#1}%
                      {\settowidth{\leftmargin}{M#1}%
                       \settowidth{\labelwidth}{#1}%
                       \settowidth{\labelsep}{M}%
                       #2%
                      }%
                }%
               {\end{list}}%
\newenvironment{EinfachListe}[1]%
               {\begin{StandardListe}{#1}{\ListNullAbstaende}}%
               {\end{StandardListe}}%
               {\begin{StandardListe}{#1}{\ListNurAnstrichAbstand}}%
               {\end{StandardListe}}%
\newcommand{\labelsatz}[1]{#1}
\newcounter{listennr}                      
\newlength{\hilfslaenge}
\newlength{\stdlabellaenge}
\newlength{\maximum}
\newcommand{\stdlabel}{}
\newcommand{\Maximum}{}
\newcommand{\iitem}[1][]{\ifthenelse{\equal{#1}{}}%
                           {\item \setlength{\hilfslaenge}{\stdlabellaenge}}%
                           {\item[\labelsatz{#1}\hfill]%
                            \settowidth{\hilfslaenge}{\labelsatz{#1}}}%
                         \ifthenelse{\lengthtest{\maximum < \hilfslaenge}}%
                           {\setlength{\maximum}{\hilfslaenge}%
                            \ifthenelse{\equal{#1}{}}%
                               {\renewcommand{\Maximum}{\stdlabel}}%
                               {\renewcommand{\Maximum}{#1}}}%
                           {}%
                      }      
\makeatletter
\newenvironment{AutoLabelLaengenListe}[2][]%
               {\begin{list}%
                      {\labelsatz{#1}\hfill}%
                      {\stepcounter{listennr}%
                       \settowidth{\leftmargin}{M\labelsatz{\ref{listnr\arabic{listennr}}}}%
                       \settowidth{\labelwidth}{\labelsatz{\ref{listnr\arabic{listennr}}}}%
                       \settowidth{\labelsep}{M}%
                       \settowidth{\stdlabellaenge}{\labelsatz{#1}}%
                       \renewcommand{\stdlabel}{#1}%
                       #2%
                       \renewcommand{\Maximum}{}%
                      }%
                }%
               {\renewcommand{\@currentlabel}{\Maximum}%
                \label{listnr\arabic{listennr}}%
                \end{list}%
                }%
\makeatother
\newenvironment{StandardEinrueckung}[2]%
               {\begin{list}%
                      {#1}%
                      {\settowidth{\leftmargin}{M#1}%
                       \settowidth{\labelwidth}{#1}%
                       \settowidth{\labelsep}{M}%
                       #2%
                      }%
                \item}%
               {\end{list}}%
\newenvironment{Einrueckungpur}[1]%
               {\begin{StandardEinrueckung}{#1}{\ListNullAbstaende}}%
               {\end{StandardEinrueckung}}%
\newenvironment{Einrueckung}[1]%
               {\begin{StandardEinrueckung}{#1}{\setlength{\parsep}{0pt}}}%
               {\end{StandardEinrueckung}}%
\newcommand{\EineZeileGleichung}[2][0.0ex]
           {
            
            \vspace{#1} 
            \noindent
            \hspace*{\fill}
            $\displaystyle{#2}$
            \hspace*{\fill}

            \vspace{#1} 
            
           }

\makeatletter
\newcommand{\EineNumZeileGleichung}[2][0.5ex]
           {
            
            \vspace{#1} 
            \noindent
            \stepcounter{equation}
            \renewcommand{\@currentlabel}{\arabic{equation}}%
            \phantom{(\arabic{equation})}\hspace*{\fill}
            $\displaystyle{#2}$
            \hspace*{\fill}
            (\arabic{equation})

            \vspace{#1} 
            
           }
\makeatother
\newcommand{\breitrel}[1]{\hspace*{\tabcolsep} #1 \hspace*{\tabcolsep}}
\newlength{\abstaug}              
\newenvironment{AllgUnnumGleichung}[2][1.0ex]
               {
  
                \setlength{\abstaug}{#1}
                \vspace{\abstaug}
                \hspace*{\fill}
                $\begin{array}[t]{#2}
                }%
               {\end{array}$
                \hspace*{\fill}
  
                \vspace{\abstaug}

                }%
               {
  
                \setlength{\abstaug}{#1}
                \vspace{\abstaug}
                $\begin{tabular*}{\textwidth}[t]{#2}
                }%
               {\end{tabular*}$

                \vspace{\abstaug}

               }%
\newcommand{\s}{\\[0ex] }             
\newenvironment{StandardUnnumGleichung}[1][0ex]
               {\renewcommand{\s}{\\[#1] }%
                \begin{AllgUnnumGleichung}{>{\displaystyle}rc>{\displaystyle}l}}%
               {\end{AllgUnnumGleichung}}%
\newcommand{\erl}[1]{\hfill\mbox{\hspace*{1.5em}\small (#1)}}

\newcommand{\erllang}[2][0.5\textwidth]%
              {\hfill\hspace*{1.5em}%
               \begin{minipage}[t]{#1}{\small%
                          \begin{list}{(}{\ListNullAbstaende%
                                          \settowidth{\leftmargin}{(}%
                                          \settowidth{\labelwidth}{(}%
                                          \settowidth{\labelsep}{}%
                                         }%
                          \item#2)%
                          \end{list}}%
               \end{minipage}\\[-0.9ex]
              }%
\newcommand{\DefBemUmgeb}[1]
           {\newenvironment{#1}[1][]%
                           {\begin{Einrueckung}{{\bf #1}}%
                            \ifx##1\empty\else{{\bf ##1}
                            
                                                        }\fi%
                            }%
                           {\end{Einrueckung}}}
\newcommand{\DefSBemUmgeb}[2]
           {\newenvironment{#1}[1][]%
                           {\begin{Einrueckung}{{\bf #2}}%
                            \ifx##1\empty\else{{\bf ##1}
                            
                                                        }\fi%
                            }%
                           {\end{Einrueckung}}}
\makeatletter
\newcommand{\DefBspUmgeb}[3]
           {\newcounter{#2}[#3]%
            \newenvironment{#1}[1][]%
                           {\stepcounter{#2}%
                            \renewcommand{\ZaehlerMarke}{\arabic{#2}}%
                            \renewcommand{\Einzugsname}{{\bf #1 \ZaehlerMarke}}%
                            \begin{Einrueckung}{\Einzugsname}
                            \ifx##1\empty\else{{\bf ##1}\\}\fi%
                            \renewcommand{\@currentlabel}{\ZaehlerMarke}%
                            }%
                           {\end{Einrueckung}}}
\makeatother
\newcommand{\ZaehlerbisEbene}{section}
\newcommand{\Ebenea}{section}
\newcommand{\Ebeneb}{subsection}

\newcommand{\Abschnittnummer}{%
            \ifx\ZaehlerbisEbene\Ebenea{\arabic{section}}%
             \else{%
              \ifx\ZaehlerbisEbene\Ebeneb{\arabic{section}.\arabic{subsection}}%
               \else{\arabic{section}.\arabic{subsection}.\arabic{subsubsection}}%
              \fi}%
            \fi}     
\newcommand{\Abschnittnummerpunkt}{\Abschnittnummer.}     
\newcommand{\Einzugsname}{}
\newcommand{\ZaehlerMarke}{}
\makeatletter
\newcommand{\DefThmUmgeb}[3]
           {\newcounter{#1}[#3]%
            \newenvironment{#1}[1][]%
                           {\stepcounter{#2}%
                            \setcounter{#1}{\value{#2}}%
                            \renewcommand{\ZaehlerMarke}{\Abschnittnummerpunkt\arabic{#1}}%
                            \renewcommand{\Einzugsname}{{\bf #1 \ZaehlerMarke}}%
                            \begin{Einrueckung}{\Einzugsname}
                            \ifx##1\empty\else{{\bf ##1}
                            
                                                        }\fi%
                            \renewcommand{\@currentlabel}{\ZaehlerMarke}%
                            }%
                           {\end{Einrueckung}}}
\makeatother
\makeatletter
\newcommand{\DefSThmUmgeb}[4]
           {\newcounter{#1}[#3]%
            \newenvironment{#1}[1][]%
                           {\stepcounter{#2}%
                            \setcounter{#1}{\value{#2}}%
                            \renewcommand{\ZaehlerMarke}{\Abschnittnummerpunkt\arabic{#1}}%
                            \renewcommand{\Einzugsname}{{\bf #4 \ZaehlerMarke}}
                            \begin{Einrueckung}{\Einzugsname}
                            \ifx##1\empty\else{{\bf ##1}

                                                        }\fi%
                            \renewcommand{\@currentlabel}{\ZaehlerMarke}%
                            }%
                           {\end{Einrueckung}}}
\makeatother
\newenvironment{Beweis}[1][]%
               {\begin{Einrueckung}{{\bf Beweis}}%
                \ifx#1\empty\else{{\bf #1}

                                            }\fi%
                }%
               {\end{Einrueckung}%
                }%
\newenvironment{Proof}[1][]%
               {\begin{Einrueckung}{{\bf Proof}}%
                \ifx#1\empty\else{{\bf #1}

                                            }\fi%
                }%
               {\end{Einrueckung}%
                }%
               {\begin{Einrueckung}{{\bf \glqq Beweis\grqq}}%
                \ifx#1\empty\else{{\bf #1}
                
                                            }\fi%
                }%
               {\end{Einrueckung}%
                }%
               {\begin{Einrueckung}{{\bf Begr"undung}}%
                \ifx#1\empty\else{{\bf #1}
                
                                            }\fi%
                }%
               {\end{Einrueckung}%
                }%
\newenvironment{Hinrichtung}%
               {\begin{Einrueckungpur}{$\impliz$}}%
               {\end{Einrueckungpur}}%
\newenvironment{Rueckrichtung}%
               {\begin{Einrueckungpur}{$\invimpliz$}}%
               {\end{Einrueckungpur}}%
               {\begin{Einrueckungpur}{\glqq$\teilmenge$\grqq}}%
               {\end{Einrueckungpur}}%
               {\begin{Einrueckungpur}{\glqq$\obermenge$\grqq}}%
               {\end{Einrueckungpur}}%
\newenvironment{SubSet}%
               {\begin{Einrueckungpur}{"$\teilmenge$"}}%
               {\end{Einrueckungpur}}%
\newenvironment{SuperSet}%
               {\begin{Einrueckungpur}{"$\obermenge$"}}%
               {\end{Einrueckungpur}}%
\newcommand{\qed}{\nopagebreak\hspace*{2em}\hspace*{\fill}{\bf qed}}
\newcommand{\ARabic}{\arabic}
\newcommand{\Nummerntypa}{\arabic}   
\newcommand{\Nummerntypb}{\alph}
\newcommand{\Nummerntypc}{\roman}
\newcommand{\Nummerntypd}{\Alph}

\newcommand{\Nra}{\Nummerntypa{Nummera}}            
\newcommand{\Nrb}{\Nummerntypb{Nummerb}}            
\newcommand{\Nrc}{\Nummerntypc{Nummerc}}                
\newcommand{\Nrd}{\Nummerntypd{Nummerd}}                
\newcommand{\ZeichenzuNrTyp}[1]%
           {\ifx#1\ARabic {.}\else{)}%
                  \fi}                              
\newcommand{\NrZeicha}{\ZeichenzuNrTyp{\Nummerntypa}}
\newcommand{\NrZeichb}{\ZeichenzuNrTyp{\Nummerntypb}}
\newcommand{\NrZeichc}{\ZeichenzuNrTyp{\Nummerntypc}}
\newcommand{\NrZeichd}{\ZeichenzuNrTyp{\Nummerntypd}}
\newcommand{\ListMarkea}%
           {\Nra\NrZeicha}
\newcommand{\ListMarkeb}%
           {\Nra\NrZeicha\Nrb\NrZeichb}
\newcommand{\ListMarkec}%
           {\Nra\NrZeicha\Nrb\NrZeichb\Nrc\NrZeichc}
\newcommand{\ListMarked}%
           {\Nra\NrZeicha\Nrb\NrZeichb\Nrc\NrZeichc\Nrd\NrZeichd}
\newcommand{\Anfangszeichen}{}
\newcommand{\Anfangspunkt}{}
\newcounter{Schachtelebene}
\newcounter{Hilfszaehler}
\newcommand{\Hilfsbefehl}{}
\newcommand{\Schachtelebene}{\alph{Schachtelebene}}
\makeatletter
\newenvironment{AllgNumerierteListe}[2][]
               {\addtocounter{Schachtelebene}{1}%
		\setcounter{Hilfszaehler}{#2}%
                \renewcommand{\Anfangszeichen}%
                             {\renewcommand{\Hilfsbefehl}{\csname Nummerntyp\Schachtelebene \endcsname}%
                              \Hilfsbefehl{Hilfszaehler}}%
                \renewcommand{\Anfangspunkt}%
                             {\csname NrZeich\Schachtelebene \endcsname}%
                \begin{list}%
                      {\stepcounter{Nummer\Schachtelebene}%
                       \csname Nr\Schachtelebene \endcsname
                       \csname NrZeich\Schachtelebene \endcsname
                       }%
                      {\settowidth{\leftmargin}{M\Anfangszeichen\Anfangspunkt}%
                       \settowidth{\labelwidth}{\Anfangszeichen\Anfangspunkt}%
                       \settowidth{\labelsep}{M}%
                       \setlength{\topsep}{0pt}%
                       \setlength{\parskip}{0pt}%
                       \setlength{\partopsep}{0pt}%
                       \setlength{\itemsep}{0pt}%
                       \setlength{\parsep}{0pt}%
                      }%
                \renewcommand{\@currentlabel}{\csname ListMarke\Schachtelebene \endcsname}%
                }%
               {\ifthenelse{\equal{}{}}{\setcounter{Nummer\Schachtelebene}{0}}{}
                \addtocounter{Schachtelebene}{-1}%
                \end{list}}
\makeatother
\newenvironment{NumerierteListe}[1]
               {\begin{AllgNumerierteListe}{#1}}
               {\end{AllgNumerierteListe}}
\newenvironment{WeiterNumerierteListe}[1]
               {\begin{AllgNumerierteListe}[Weiter]{#1}}
               {\end{AllgNumerierteListe}}

\newcommand{\UnnumAnfangszeichen}{}
\newcounter{UnnumSchachtelebene}
\newcommand{\UnnumSchachtelebene}{\alph{UnnumSchachtelebene}}
\makeatletter
\newenvironment{UnnumerierteListe}%
               {\addtocounter{UnnumSchachtelebene}{1}%
                \renewcommand{\UnnumAnfangszeichen}%
                             {\csname UnnumZeich\UnnumSchachtelebene \endcsname}%
                \begin{list}%
                      {\UnnumAnfangszeichen}%
                      {\settowidth{\leftmargin}{M\UnnumAnfangszeichen}%
                       \settowidth{\labelwidth}{\UnnumAnfangszeichen}%
                       \settowidth{\labelsep}{M}%
                       \setlength{\topsep}{0pt}%
                       \setlength{\parskip}{0pt}%
                       \setlength{\partopsep}{0pt}%
                       \setlength{\itemsep}{0pt}%
                       \setlength{\parsep}{0pt}%
                      }%
                }%
               {\addtocounter{UnnumSchachtelebene}{-1}%
                \end{list}}
\makeatother
\newlength{\fktdefhilfslaenge}
\newcommand{\fktdef}[5]
           {\hspace*{\fill}
            $\begin{array}[t]{cccc}%
            #1: & #2 & \nach & #3 \\    
                & #4 & \auf  & #5
            \end{array}$
            \settowidth{\fktdefhilfslaenge}{$#1$:}
            \hspace*{0.6 \fktdefhilfslaenge}  
            \hspace*{\fill}}
\newcommand{\fktdefpur}[5]
           {$\begin{array}[t]{cccc}%
            #1: & #2 & \nach & #3 \\    
                & #4 & \auf  & #5
            \end{array}$}
\newcommand{\fktdefabgesetzt}[5]
           {
           
            \hspace*{\fill}
            $\begin{array}[t]{cccc}%
            #1: & #2 & \nach & #3 \\    
                & #4 & \auf  & #5
            \end{array}$
            \settowidth{\fktdefhilfslaenge}{$#1$:}
            \hspace*{0.6 \fktdefhilfslaenge}  
            \hspace*{\fill}
            
            }
\newcommand{\sectioninh}[1]%
           {\section*{#1}%
            \addcontentsline{toc}{section}{#1}}
\newcommand{\anhang}%
           {\appendix
            \sectioninh{Anhang}
            \renewcommand{\Abschnittnummer}{%
                  \ifx\ZaehlerbisEbene\Ebenea{\Alph{section}}%
                  \else{%
                        \ifx\ZaehlerbisEbene\Ebeneb{\Alph{section}.\arabic{subsection}}%
                        \else{\Alph{section}.\arabic{subsection}.\arabic{subsubsection}}%
                        \fi}%
                  \fi}%
            \renewcommand{\Abschnittnummerpunkt}{\Abschnittnummer.}     
            }            
\newcommand{\anhangengl}%
           {\appendix
            \sectioninh{Appendix}
            \renewcommand{\Abschnittnummer}{%
                  \ifx\ZaehlerbisEbene\Ebenea{\Alph{section}}%
                  \else{%
                        \ifx\ZaehlerbisEbene\Ebeneb{\Alph{section}.\arabic{subsection}}%
                        \else{\Alph{section}.\arabic{subsection}.\arabic{subsubsection}}%
                        \fi}%
                  \fi}%
            \renewcommand{\Abschnittnummerpunkt}{\Abschnittnummer.}     
            }            
\newlength{\querfhilfsl}              

\newlength{\hll}

\newcommand{\keinseitenumbr}{\nopagebreak[4]}

\makeatletter
\newcommand{\EineErwNumZeileGleichung}[2][0.5ex]
           {
            
            \vspace{#1} 
            \noindent
            \stepcounter{equation}
            \renewcommand{\@currentlabel}{\arabic{equation}}%
            \phantom{(\arabic{equation})}\hspace*{\fill}
            #2 
            \hspace*{\fill}
            (\arabic{equation})

            \vspace{#1} 
            
           }
\makeatother
\DefThmUmgeb{Theorem}{Theorem}{\ZaehlerbisEbene}
\DefThmUmgeb{Definition}{Definition}{\ZaehlerbisEbene}
\DefThmUmgeb{Satz}{Theorem}{\ZaehlerbisEbene}
\DefThmUmgeb{Proposition}{Theorem}{\ZaehlerbisEbene}
\DefThmUmgeb{Lemma}{Theorem}{\ZaehlerbisEbene}
\DefThmUmgeb{Folgerung}{Theorem}{\ZaehlerbisEbene}
\DefThmUmgeb{Corollary}{Theorem}{\ZaehlerbisEbene}
\DefThmUmgeb{Vorschrift}{Definition}{\ZaehlerbisEbene}
\DefThmUmgeb{Construction}{Definition}{\ZaehlerbisEbene}
\DefSThmUmgeb{FormSatz}{Theorem}{\ZaehlerbisEbene}{\glqq Satz\grqq} 
\DefThmUmgeb{Vermutung}{Theorem}{\ZaehlerbisEbene}
\DefBspUmgeb{Beispiel}{Beispiel}{subsubsection}
\DefBemUmgeb{Bemerkung}
\DefBemUmgeb{Remark}
\DefSBemUmgeb{OffeneFrage}{Offene Frage}
\newcommand{\bdf}{\begin{Definition}}
\newcommand{\edf}{\end{Definition}}
\newcommand{\bvorsch}{\begin{Vorschrift}}
\newcommand{\evorsch}{\end{Vorschrift}}
\newcommand{\bconst}{\begin{Construction}}
\newcommand{\econst}{\end{Construction}}
\newcommand{\bthm}{\begin{Theorem}}
\newcommand{\ethm}{\end{Theorem}}
\newcommand{\bsatz}{\begin{Satz}}
\newcommand{\esatz}{\end{Satz}}
\newcommand{\bprop}{\begin{Proposition}}
\newcommand{\eprop}{\end{Proposition}}
\newcommand{\blem}{\begin{Lemma}}
\newcommand{\elem}{\end{Lemma}}
\newcommand{\bfolg}{\begin{Folgerung}}
\newcommand{\efolg}{\end{Folgerung}}
\newcommand{\bcorr}{\begin{Corollary}}
\newcommand{\ecorr}{\end{Corollary}}
\newcommand{\bbew}{\begin{Beweis}}
\newcommand{\ebew}{\end{Beweis}}
\newcommand{\bpf}{\begin{Proof}}
\newcommand{\epf}{\end{Proof}}
\newcommand{\bwnum}{\begin{WeiterNumerierteListe}}
\newcommand{\ewnum}{\end{WeiterNumerierteListe}}
\newcommand{\bbem}{\begin{Bemerkung}}
\newcommand{\ebem}{\end{Bemerkung}}
\newcommand{\brem}{\begin{Remark}}
\newcommand{\erem}{\end{Remark}}
\newcommand{\bnum}{\begin{NumerierteListe}}
\newcommand{\enum}{\end{NumerierteListe}}
\newcommand{\bunum}{\begin{UnnumerierteListe}}
\newcommand{\eunum}{\end{UnnumerierteListe}}
\newcommand{\bbsp}{\begin{Beispiel}}
\newcommand{\ebsp}{\end{Beispiel}}
\newcommand{\bof}{\begin{OffeneFrage}}
\newcommand{\eof}{\end{OffeneFrage}}
\newcommand{\bgl}{\begin{StandardUnnumGleichung}}
\newcommand{\egl}{\end{StandardUnnumGleichung}}
\newcommand{\zgl}{\EineZeileGleichung}

\newcommand{\znumgl}{\EineNumZeileGleichung}

\newcommand{\berlgl}{\begin{StandardUnnumGleichung}}
\newcommand{\eerlgl}{\end{StandardUnnumGleichung}}
\newcommand{\beinrueck}{\begin{Einrueckungpur}} 
\newcommand{\eeinrueck}{\end{Einrueckungpur}}
\newcommand{\beinflist}{\begin{EinfachListe}} 
\newcommand{\eeinflist}{\end{EinfachListe}}
\newcommand{\beq}{\begin{equation}}
\newcommand{\eeq}{\end{equation}}
\newcommand{\bhin}{\begin{Hinrichtung}}
\newcommand{\ehin}{\end{Hinrichtung}}
\newcommand{\brueck}{\begin{Rueckrichtung}}
\newcommand{\erueck}{\end{Rueckrichtung}}
\newcommand{\bvl}{\begin{AutoLabelLaengenListe}{\ListNullAbstaende}}
\newcommand{\evl}{\end{AutoLabelLaengenListe}}
\newcommand{\df}[1]{{\bf #1}}

\begin{document}
\title{Stratification of the Generalized Gauge Orbit Space}
\author{Christian Fleischhack\thanks{e-mail: 
            Christian.Fleischhack@itp.uni-leipzig.de {\it or}    
            Christian.Fleischhack@mis.mpg.de} \\   
        \\
        \begin{minipage}{0.43\textwidth}
        \begin{center}
        {\normalsize\em Mathematisches Institut}\\[\adressabstand]
        {\normalsize\em Universit\"at Leipzig}\\[\adressabstand]
        {\normalsize\em Augustusplatz 10/11}\\[\adressabstand]
        {\normalsize\em 04109 Leipzig, Germany}\\
        \end{center}
        \end{minipage}
        \begin{minipage}{0.43\textwidth}
        \begin{center}
        {\normalsize\em Institut f\"ur Theoretische Physik}\\[\adressabstand]
        {\normalsize\em Universit\"at Leipzig}\\[\adressabstand]
        {\normalsize\em Augustusplatz 10/11}\\[\adressabstand]
        {\normalsize\em 04109 Leipzig, Germany}\\
        \end{center}
        \end{minipage} \\[-10\adressabstand]
        {\normalsize\em Max-Planck-Institut f\"ur Mathematik in den
                        Naturwissenschaften}\\[\adressabstand]
        {\normalsize\em Inselstra\ss e 22-26}\\[\adressabstand]
        {\normalsize\em 04103 Leipzig, Germany}}
\date{January 5, 2000}
\maketitle
\begin{abstract}
The action of Ashtekar's generalized gauge group $\Gb$ on the space $\Ab$
of generalized connections is investigated for compact structure groups $\LG$.

First a stratum is defined to be the set of all connections of 
one and the same gauge orbit type, 
i.e. the conjugacy class of the centralizer of the holonomy group. 
Then a slice theorem is proven on $\Ab$. This yields the openness of 
the strata. Afterwards, a denseness theorem is proven for the strata. 
Hence, $\Ab$ is topologically regularly stratified 
by $\Gb$. These results coincide with those of Kondracki and Rogulski 
for Sobolev connections. As a by-product, we prove that the set of all
gauge orbit types equals the set of all 
(conjugacy classes of) Howe subgroups
of $\LG$.
Finally, we show that the set of all gauge orbits with maximal type
has the full induced Haar measure $1$.
\end{abstract}
\newpage
\section{Introduction}
For quite a long time the geometric structure of gauge theories has been
investigated. A classical (pure) gauge theory consists of three basic objects:
First the set $\A$ of smooth connections ("gauge fields")
in a principle fiber bundle, 
then the set $\G$ of all smooth gauge transforms, i.e. automorphisms of this
bundle, and finally the action of $\G$ on $\A$.
Physically, two gauge fields that are related by a gauge transform 
describe one and the same situation. Thus, the
space of all gauge orbits, i.e. elements in $\ag$, is the configuration
space for the gauge theory. Unfortunately, in contrast to $\Ab$, which is
an affine space, the space $\ag$ has a very complicated structure: It
is non-affin, non-compact and infinite-dimensional and it is not 
a manifold. This causes enormous problems, in particular, when one wants
to quantize a gauge theory. One possible quantization method is
the path integral quantization. Here one has to find an appropriate measure
on the configuration space of the classical theory, hence a measure
on $\ag$. As just indicated, this is very hard to find. Thus, one
has hoped for a better understanding of the structure of $\ag$.
However, up to now, results are quite rare. 

About 20 years ago, the efforts were focussed on a related problem:
The consideration of connections and gauge transforms that
are contained in a certain Sobolev class (see, e.g., \cite{f1}). 
Now, $\G$ is a Hilbert-Lie group and acts smoothly on $\A$.
About 15 years ago, Kondracki and Rogulski \cite{f11} 
found lots of fundamental properties of this action.
Perhaps, the most remarkable theorem they obtained 
was a slice theorem on $\A$. This means, for every orbit 
$A\circ\G\teilmenge\A$ there is an equivariant retraction from
a (so-called tubular) neighborhood of $A$ onto $A\circ\G$.
Using this theorem they could clarify the structure of the so-called
strata. A stratum contains all connections that have
the same, fixed type, i.e. the same (conjugacy class of the) stabilizer 
under the action of $\G$. Using a denseness theorem for the strata,
Kondracki and Rogulski proved that
the space $\A$ is regularly stratified
by the action of $\G$. In particular, all the strata are smooth submanifolds
of $\A$.

Despite these results the mathematically rigorous construction of a measure 
on $\ag$ has not been achieved. This problem was solved -- at least preliminary --
by Ashtekar et al. \cite{a72,a48}, but, however, not for $\ag$ itself.
Their idea was to drop simply all smoothness conditions for the connections
and gauge transforms. In detail, they first used the fact that a
connection can always be reconstructed uniquely by its
parallel transports. On the other hand, these parallel transports
can be identified with an assignment of elements
of the structure group $\LG$
to the paths in the base manifold $M$ such that 
the concatenation of paths corresponds to the product of these
group elements. It is intuitively clear that for
smooth connections the parallel transports additionally depend
smoothly on the paths \cite{d20}. But now this restriction
is removed for the generalized connections. They are only 
homomorphisms from the groupoid $\Pf$ of paths to the 
structure group $\LG$. Analogously, the set $\Gb$ of generalized
gauge transforms collects all functions from $M$ to $\LG$. Now the
action of $\Gb$ to $\Ab$ is defined purely algebraically.
Given $\Ab$ and $\Gb$ the topologies induced by the topology of $\LG$,
one sees that, for compact $\LG$, these spaces are again compact.
This guarantees the existence of a natural induced Haar measure on $\Ab$ and
$\AbGb$, the new configuration space for the path integral quantization.

Both from the mathematical and from the physical point of view it
is very interesting how the "classical" regular gauge theories are 
related to the generalized formulation in the Ashtekar framework.
First of all, it has been proven that $\A$ and $\G$ are dense subsets
in $\Ab$ and $\Gb$, respectively \cite{e8}. 
Furthermore, $\A$ is contained in a set of induced Haar measure zero \cite{a42}.
These properties coincide exactly
with the experiences known from the Wiener or Feynman path integral.
Then the Wilson loop expectation values have been determined for the
two-dimensional pure Yang-Mills theory \cite{a6,paper1} -- in coincidence
with the known results in the standard framework.
In the present paper we continue the investigations on how the results of
Kondracki and Rogulski can be extended to the Ashtekar framework.
In a previous paper \cite{paper2} we have already shown that 
the gauge orbit type is determined by the centralizer of the holonomy group.
This closely related to the observations of Kondracki and Sadowski 
\cite{f12}.
In the present paper we are going to prove that there is a slice theorem
and a denseness theorem for the space of connections in the Ashtekar framework as
well. However, our methods are completely different to those of
Kondracki and Rogulski.

\leerezeile

The outline of the paper is as follows:

After fixing the notations we prove a very crucial lemma in section
\ref{abschn:finitelem(centr)}: Every centralizer in a compact Lie group
is finitely generated. This implies that every orbit type (being
the centralizer of the holonomy group) is determined
by a finite set of holonomies of the corresponding connection.

Using the projection onto these holonomies we can lift the slice theorem
from an appropriate finite-dimensional $\LG^n$ to the space $\Ab$. This is
proven in section \ref{abschn:slice-thm_Ab} and it implies the openness of the
strata as shown in the following section.

Afterwards, we prove a denseness theorem for the strata. For this we need
a construction for new connections from \cite{paper3}. 
As a corollary we obtain that the set of all gauge orbit types equals
the set of all conjugacy classes of Howe subgroups of $\LG$. A Howe subgroup 
is a subgroup that is the centralizer of some subset of $\LG$.
This way we completely determine all possible gauge orbit types. This
has been succeeded for the Sobolev connections 
-- to the best of our knowlegde -- only for $\LG=SU(n)$ and low-dimensional $M$ 
\cite{f14}.

In Section \ref{abschn:stratifiz} we show that 
the slice and the denseness theorem yield again a 
topologically regular stratification of $\Ab$ as well as of $\AbGb$.
But, in contrast to the Sobolev case, the strata are not proved to be
manifolds.

Finally, we show in Section \ref{section:noncomplconn}
that the generic stratum (it collects the connections of
maximal type) is not only dense in $\Ab$, but has also the total induced
Haar measure $1$. This shows that the Faddeev-Popov determinant 
for the projection $\Ab\nach\AbGb$ is equal to $1$.

\section{Preliminaries}
As we indicated in \cite{paper2} the present paper is the final one
in a small series of three papers. 
In the first one \cite{paper2} we extended the definitions and 
propositions for $\Ab$, $\Gb$ and $\AbGb$ made by Ashtekar et al. 
from the case of graphs \cite{a72,a48,a30,a28,a42} and of 
webs \cite{d3} to arbitrarily smooth paths. Moreover, in that paper
we determined the gauge orbit type of a connection.
In the second paper \cite{paper3} we investigated properties of
$\Ab$ and proved, in particular, the existence of an Ashtekar-Lewandowski
measure in our context.
Now, we summarize the most important notations, definitions and facts used in
the following. For detailed information we 
refer the reader to the preceding papers \cite{paper2,paper3}.

\bunum
\item
Let $\LG$ be a {\em compact} Lie group.
\item 
A path (usually denoted by $\gamma$ or $\delta$) is a piecewise
$C^r$-map from $[0,1]$ into a connected $C^r$-manifold $M$, $\dim M\geq 2$,
$r\in\N^+\cup\{\infty\}\cup\{\omega\}$ arbitrary, but fixed. Additionally,
we fix now the decision whether we restrict the paths to be
piecewise immersive or not. Paths
can be multiplied as usual by concatenation. A graph is a finite union of 
paths, such that different paths intersect each other at most in
their end points. Paths in a graph are called simple. A path
is called finite iff it is up to the parametrization a finite
product of simple paths. Two paths are equivalent iff
the first one can be reconstructed from the second one by a sequence
of reparametrizations or of insertions or deletions of retracings.
We will only consider equivalence classes of finite paths and 
graphs. The set of (classes of) paths is denoted by $\Pf$, that of paths
from $x$ to $y$ by $\pf{xy}$ and that of
loops (paths with a fixed initial and terminal point $m$) by $\hg$,
the so-called hoop group.
\item
A generalized connection $\qa\in\Ab$ is a 
homomorphism\footnote{Homomorphism
means $h_\qa(\gamma_1\gamma_2) = h_\qa(\gamma_1) h_\qa(\gamma_2)$
supposed $\gamma_1\gamma_2$ is defined.} 
$h_\qa:\Pf\nach\LG$. (We usually write $h_\qa$ synonymously for $\qa$.)
A generalized gauge transform $\qg\in\Gb$ is a map
$\qg:M\nach\LG$. The value $\qg(x)$ of the gauge transform
in the point $x$ is usually denoted by $g_x$.
The action of $\Gb$ on $\Ab$ is given
by 
\znumgl{h_{\qa\circ\qg}(\gamma) := 
        g^{-1}_{\gamma(0)} \: h_\qa(\gamma) \: g_{\gamma(1)} 
        \text{ for all } \gamma\in\Pf. \label{action}} 
We have $\AbGb \iso \Hom(\hg,\LG)/\Ad$.
\item
Now, let $\GR$ be a graph with $\Edg(\GR)=\{e_1,\ldots,e_E\}$ 
being the set of edges and $\Ver(\GR)=\{v_1,\ldots,v_V\}$ 
the set of vertices. 
The projections onto the lattice gauge theories are defined by

\fktdef{\pi_\GR}{\Ab}{\Ab_\GR\ident\LG^E}
                {\qa}{\bigl(h_\qa(e_1),\ldots,h_\qa(e_E)\bigr)}
and
\fktdef{\pi_\GR}{\Gb}{\Gb_\GR\ident\LG^V.}
                {\qg}{\bigl(g_{v_1},\ldots,g_{v_V}\bigr)}

The topologies on $\Ab$ and $\Gb$ are the topologies generated 
by these projections. Using these topologies the action
$\Theta:\Ab\kreuz\Gb\nach\Ab$ defined by \eqref{action} 
is continuous.
Since $\LG$ is compact Lie, $\Ab$ and $\Gb$ are compact 
Hausdorff spaces and consequently completely regular.
\item
The holonomy group $\holgr_\qa$ of a connection $\qa$ is defined
by $\holgr_\qa := h_\qa(\hg)\teilmenge\LG$, its centralizer is denoted
by $Z(\holgr_\qa)$. The stabilizer of a connection $\qa\in\Ab$ 
under the action of $\Gb$ is denoted by $\bz(\qa)$. 
We have $\qg\in\bz(\qa)$ iff $g_m\in Z(\holgr_\qa)$
and for all $x\in M$ there is a path $\gamma\in\pf{mx}$
with $h_\qa(\gamma) = g_m^{-1} h_\qa(\gamma) g_x$.
In \cite{paper2} we proved that $\bz(\qa)$ and $Z(\holgr_\qa)$
are homeomorphic.
\item
The type of a gauge orbit $\eo_\qa:=\qa\circ\Gb$ is
the centralizer of the holonomy group of $\qa$ modulo 
conjugation in $\LG$. (An equivalent definition uses the stabilizer
$\bz(\qa)$ itself.)
\eunum

\section{Partial Ordering of Types}
\bdf
A subgroup $U$ of $\LG$ is called \df{Howe subgroup} iff there is a set
$V\teilmenge\LG$ with $U = Z(V)$.
\edf
Analogously to the general theory we define
a partial ordering for the gauge orbit types \cite{BourbakiLie9Russ}.
\bdf
Let $\T$ denote the set of all Howe subgroups of $\LG$.

Let $t_1, t_2\in\T$. Then $t_1 \leq t_2$ holds iff there are
$\LG_1\in t_1$ and $\LG_2\in t_2$ with $\LG_1\obermenge\LG_2$.
\edf
Obviously, we have 
\blem
The maximal element 
in $\T$ is the class $t_{\max}$ of the center $Z(\LG)$ of $\LG$,
the minimal is the class $t_{\min}$ of $\LG$ itself.
\elem
\bdf
Let $t\in\T$. We define the following expressions:
\bgl
  \Abgeq{t} & := & \{\qa\in\Ab\mid\typ(\qa) \geq t\}\s
  \Abeq{t}  & := & \{\qa\in\Ab\mid\typ(\qa) = t\}\s
  \Ableq{t} & := & \{\qa\in\Ab\mid\typ(\qa) \leq t\}.
\egl
All the $\Abeq t$ are called \df{strata}.\footnote{The justification
for that notation can be found in section \ref{abschn:stratifiz}.}
\edf

\section{Reducing the Problem to Finite-Dimensional $\LG$-Spaces}
\label{abschn:reduktion}
\label{abschn:finitelem(centr)}
\subsection{Finiteness Lemma for Centralizers}
We start with the crucial
\blem
\label{exkelemzentr}
Let $U$ be a subset of a compact Lie group $\LG$. 
Then there exist an $n\in\N$ and
$u_1,\ldots,u_n\in U$, such that $Z(\{u_1,\ldots,u_n\}) = Z(U)$.
\elem
\bpf
\bunum
\item
The case $Z(U) = \LG = Z(\leeremenge)$ is trivial.
\item
Let $Z(U)\neq\LG$. Then there is a $u_1\in U$ with $Z(\{u_1\}) \neq \LG$.
Choose now for $i\geq 1$ successively $u_{i+1}\in U$ with
$Z(\{u_1,\ldots,u_i\}) \echteobermenge Z(\{u_1,\ldots,u_{i+1}\})$ as long
as there is such a $u_{i+1}$. This procedure stops after a finite
number of steps, since each non-increasing sequence of compact 
subgroups in $\LG$ stabilizes \cite{BourbakiLie9Russ}.  
(Centralizers are always closed, thus compact.)
Therefore there is an $n\in\N$, such that
$Z(\{u_1,\ldots,u_n\}) = Z(\{u_1,\ldots,u_n\}\cup\{u\})$ for all $u\in U$. 
Thus, we have
$Z(\{u_1,\ldots,u_n\}) 
   = \bigcap_{u\in U} Z(\{u_1,\ldots,u_n\}\cup\{u\})
   = Z(\{u_1,\ldots,u_n\}\cup U)
   = Z(U)$.
\qed
\eunum
\epf   
\bcorr
\label{endlbestzentr}
Let $\qa\in\Ab$. 

Then there is a finite set $\ga\teilmenge\hg$, such that
$Z(\holgr_\qa) = Z(h_\qa(\ga))$.\footnote{$h_\qa(\ga)
:= \bigl\{h_{\qa}(\alpha_1),\ldots,h_{\qa}(\alpha_n)\bigr\}\teilmenge\LG$ 
where $n:=\elanz\ga$.
To avoid cumbersome notations we denote also
$\bigl(h_{\qa}(\alpha_1),\ldots,h_{\qa}(\alpha_n)\bigr)\in\LG^n$ by
$h_\qa(\ga)$. It should be clear from the context what is meant.
Furthermore, $\ga$ is always finite.}
\ecorr
\bpf
Due to $\holgr_{\qa}\teilmenge\LG$ and the just proven lemma there
are an $n\in\N$ and $g_1,\ldots, g_n\in\holgr_{\qa}$ with 
$Z(\{g_1,\ldots,g_n\}) = Z(\holgr_{\qa})$.
On the other hand, since $g_1,\ldots, g_n\in\holgr_{\qa}$, there are 
$\alpha_1,\ldots,\alpha_n\in\hg$ with $g_i = h_{\qa}(\alpha_i)$ for all
$i=1,\ldots,n$.
\qed
\epf
\subsection{Reduction Mapping}
\bdf
Let $\ga\teilmenge\hg$. Then the map 
\fktdefabgesetzt{\varphi_\ga}{\Ab}{\LG^{\elanz\ga}}
                {\qa}{h_{\qa}(\ga)}
is called \df{reduction mapping}.
\edf
\blem
Let $\ga\teilmenge\hg$ be arbitrary.

Then $\varphi_\ga$ is continuous, and
for all $\qa\in\Ab$ and $\qg\in\Gb$ we have
$\varphi_\ga(\qa\circ\qg) = \varphi_\ga(\qa)\circ g_m$.
Here $\LG$ acts on $\LG^{\elanz\ga}$ by the adjoint map.
\elem
\bpf
\bunum
\item
$\varphi_\ga:\Ab\nach\LG^{\elanz\ga}$ is as a map into a product space
continuous iff $\pi_i\circ\varphi_\ga\ident\varphi_{\{\alpha_i\}}$ is
continuous
for all projections $\pi_i:\LG^{\elanz\ga}\nach\LG$ onto the $i$th factor.
Thus, it is sufficient to prove the continuity of $\varphi_{\{\alpha\}}$
for all $\alpha\in\hg$.

Now decompose $\alpha$ into a product of finitely many edges $e_j$, 
$j=1,\ldots,J$ (i.e., into paths that can be represented as an edge
in a graph).
Then the mapping $\Ab\nach\LG^J$ with
$\qa\auf\bigl(\pi_{e_1}(\qa),\ldots,\pi_{e_J}(\qa)\bigr)$ is continuous
per definitionem. Since the multiplication in $\LG$ is continuous,
$\varphi_{\{\alpha\}}$ is continuous, too.
\item
The compatibility with the group action follows from 
$h_{\qa\circ\qg}(\ga) = g_m^{-1}\:h_\qa(\ga)\:g_m$.
\qed
\eunum
\epf
\subsection{Adjoint Action of $\LG$ on $\LG^n$}
In this short subsection we will summarize the most important facts about
the adjoint action of $\LG$ on $\LG^n$ that can be
deduced from the general theory of transformation groups
(see, e.g., \cite{Bredon}).

First we determine the stabilizer $\LG_{\vec g}$ of an element $\vec g\in\LG^n$.
We have
\zgl{
 \LG_{\vec g} = \{g\in\LG\mid \vec g\circ g = \vec g\}
              = \{g\in\LG\mid g^{-1} g_i g = g_i \fueralle i\}
              = Z(\{g_1,\ldots,g_n\}).}
Consequently, we have for the type of the corresponding orbit
\zgl{\typ(\vect g) = [\LG_{\vect g}] 
                   = [Z(\{g_1,\ldots,g_n\})].}
The slice theorem reads now as follows:
\bprop
\label{Slicethm_auf_LG^n}
Let $\vec g\in\LG^n$. Then there is an
$S\teilmenge\LG^n$ with $\vec g\in S$, such that:
\bunum
\item
$S\circ\LG$ is an open neighboorhood of $\vec g\circ\LG$ and
\item
there is an equivariant retraction $f:S\circ \LG\nach \vec g\circ\LG$ with
$f^{-1}(\{\vec g\}) = S$.
\eunum
\eprop
Both on $\Ab$ and on $\LG^n$ the type is a Howe subgroup
of $\LG$. The transformation behaviour
of the types under a reduction mapping
is stated in the next
\bprop
Any reduction mapping is type-minorifying, i.e. 
for all $\ga\teilmenge\hg$ and all $\qa\in\Ab$ 
we have \zgl{\typ\bigl(\varphi_\ga(\qa)\bigr) \leq \typ(\qa).}
\eprop
\bpf
We have
$\typ\bigl(\varphi_\ga(\qa)\bigr) = [Z(\varphi_\ga(\qa))]
                             \ident [Z(h_\qa(\ga))]
                               \leq [Z(\holgr_\qa)]
                                  = \typ(\qa).$
\qed
\epf
\section{Slice Theorem for $\Ab$}
\label{abschn:slice-thm_Ab}
We state now the main theorem of the present paper.
\bthm
There is a tubular neighbourhood for any gauge orbit.

Equivalently we have: For all $\qa\in\Ab$ there is an
$\qS\teilmenge\Ab$ with $\qa\in\qS$, such that:
\bunum
\item
$\qS\circ\Gb$ is an open neighbourhood of $\qa\circ\Gb$ and
\item
there is an equivariant retraction $F:\qS\circ\Gb\nach\qa\circ\Gb$ with
$F^{-1}(\{\qa\})=\qS$.
\eunum
\ethm
\subsection{The Idea}
Our proof imitates in a certain sense 
the proof of the standard slice theorem (see, e.g., \cite{Bredon})
which is valid for the action 
of a {\em finite-dimensional} compact {\em Lie} group $G$ on
a Hausdorff space $X$.
Let us review the main idea of this proof.
Given $x\in X$. Let $H\teilmenge G$ be the stabilizer of $x$,
i.e., $[H]$ is an orbit type on the $G$-space $X$.
Now, this situation is simulated on an $\R^n$, i.e., for an appropriate
action of $G$ on $\R^n$ one chooses a point with stabilizer $H$. 
So the orbits on $X$ and on $\R^n$ can be identified.
For the case of $\R^n$ the proof of a slice theorem is not very
complicated. The crucial point of the general proof is the usage of
the Tietze-Gleason extension theorem because this yields an
equivariant extension $\psi:X\nach\R^n$, mapping one orbit
onto the other. Finally, by means of $\psi$ the slice theorem
can be lifted from $\R^n$ to $X$.

What can we learn for our problem? Obviously, $\Gb$ is not
a finite-dimensional Lie group. But, we know that the stabilizer
$\bz(\qa)$ of a connection is homeomorphic to the centralizer
$Z(\holgr_\qa)$ of the holonomy group that is a subgroup of $\LG$.
Since every centralizer is finitely generated,
$Z(\holgr_\qa)$ equals $Z(h_\qa(\ga))$
with an appropriate finite $\ga\in\hg$. This is nothing but the 
stabilizer of the adjoint action of $\LG$ on $\LG^n$. 
Thus, the reduction mapping $\varphi_\ga$ is the desired equivalent for
$\psi$.

We are now looking for an appropriate $\qS\teilmenge\Ab$, such that
\fktdefabgesetzt{F}{\qS\circ\Gb}{\qa\circ\Gb}{\qa'\circ\qg}{\qa\circ\qg}

is well-defined and has the desired properties.

In order to make $F$ well-defined, we need 
$\qa'\circ \qg = \qa'$ $\impliz$ $\qa\circ \qg = \qa$
for all $\qa'\in\qS$ and $\qg\in\Gb$, i.e.
$\bz(\qa') \teilmenge \bz(\qa)$.
Applying the projections $\pi_x$ on the stabilizers (see \cite{paper2})
we get for $\gamma_x\in\pf{mx}$ (let $\gamma_m$ be the 
trivial path)
\zgl{h_{\qa'}(\gamma_m)^{-1} Z(\holgr_{\qa'}) h_{\qa'}(\gamma_x) 
      = \pi_x(\bz(\qa')) \teilmenge \pi_x(\bz(\qa)) =
     h_{\qa}(\gamma_m)^{-1} Z(\holgr_{\qa}) h_{\qa}(\gamma_x),}
thus     
\znumgl{\label{slicethm_gl} Z(\holgr_{\qa'}) \teilmenge
        h_{\qa'}(\gamma_m) h_{\qa}(\gamma_m)^{-1} \: Z(\holgr_{\qa}) \:
        h_{\qa}(\gamma_x) h_{\qa'}^{-1}(\gamma_x)}
for all $x\in M$. In particular, we have 
$Z(\holgr_{\qa'}) \teilmenge Z(\holgr_{\qa})$ for $x=m$.

Now we choose an $\ga\teilmenge\hg$ with $Z(\holgr_\qa) = Z(h_\qa(\ga))$
and an $S\teilmenge \LG^{\elanz\ga}$
and an equivariant retraction 
$f:S\circ \LG \nach \varphi_\ga(\qa)\circ \LG$.
Since equivariant mappings magnify stabilizers (or at least do not reduce
them), we have $Z(\vec g')\teilmenge Z(\varphi_\ga(\qa))$ for all
$\vec g'\in S$.

Therefore, the condition of \eqref{slicethm_gl} would be, e.g., 
fulfilled if we had for all $\qa'\in\qS$
\bnum{2}
\item
$\varphi_\ga(\qa') \in S$ and
\item
$h_{\qa'}(\gamma_x) = h_{\qa}(\gamma_x)$ for all $x\in M$,
\enum
because the first condition
implies $Z(\holgr_{\qa'}) \teilmenge Z(h_{\qa'}(\ga)) 
                          \ident Z(\varphi_\ga(\qa'))
                          \teilmenge Z(\varphi_\ga(\qa)) = Z(\holgr_\qa)$.
We could now choose $\qS$ such that these two conditions are fulfilled.
However, this would imply $F^{-1}(\{A\}) \echteobermenge \qS$ in general
because for $\qg\in\bz(\qa)$ together with $\qa'$ the 
connection $\qa'\circ \qg$ 
is contained in $F^{-1}(\{A\})$ as well,\footnote{We have
$F(\qa') = \qa = \qa\circ\qg = F(\qa'\circ\qg)$.} but $\qa'\circ\qg$
needs no longer fulfill the two conditions above.
Now it is quite obvious to define $\qS$ as the set of all connections
fulfilling these conditions multiplied with $\bz(\qa)$.
And indeed, the well-definedness remains valid.

\subsection{The Proof}
\bpf
\bnum{10}
\item
Let $\qa\in\Ab$. Choose for $\qa$ an $\ga\teilmenge\hg$ with
$Z(\holgr_\qa) = Z(h_\qa(\ga))$
according to Corollary \ref{endlbestzentr}
and denote the corresponding reduction mapping 
$\varphi_\ga:\Ab\nach\LG^{\elanz\ga}$ shortly by $\varphi$.
\item
Due to Proposition \ref{Slicethm_auf_LG^n} there is an
$S\teilmenge \LG^{\elanz\ga}$ with $\varphi(\qa)\in S$, such that
\bunum
\item
$S\circ\LG$ is an open neighbourhood of $\varphi(\qa)\circ\LG$ and
\item
there exists an equivariant mapping $f$ with
\bunum
\item
$f:S\circ \LG\nach \varphi(\qa)\circ\LG$ and
\item
$f^{-1}(\{\varphi(\qa)\}) = S$.
\eunum
\eunum
\item
We define the mapping
\fktdefabgesetzt{\psi}{\Ab}{\Gb,}{\qa'}{\bigl(h_{\qa'}(\gamma_x)\bigr)_{x\in M}}
whereas for all $x\in M\setminus\{m\}$ the (arbitrary, but fixed)
path $\gamma_x$ runs from $m$ to
$x$ and $\gamma_m$ is the trivial path.
\item
As we motivated above we set 
\bgl[1ex]
\qS_0 & := & \phantom{\bigl(}\varphi^{-1}(S) \: \cap \: \psi^{-1}(\psi(\qa)), \s
\qS\phantom{{}_0} & := & \bigl(\varphi^{-1}(S) \: \cap \: \psi^{-1}(\psi(\qa))\bigr) \circ
           \bz(\qa) \breitrel\ident \qS_0\circ\bz(\qa)
\egl
and
\fktdefabgesetzt{F}{\qS\circ\Gb}{\qa\circ\Gb.}{\qa'\circ\qg}{\qa\circ\qg}
\item
$F$ is well-defined.
\bunum
\item
Let $\qa'\circ\qg' = \qa''\circ\qg''$ with
$\qa',\qa''\in\qS$ and $\qg',\qg''\in\Gb$.
Then there exist $\qz',\qz''\in\bz(\qa)$ with $\qa'=\qa_0'\circ\qz'$ and 
$\qa''=\qa_0''\circ\qz''$ as well as $\qa'_0,\qa''_0\in \qS_0$.
\item
Due to $\qS_0\teilmenge\psi^{-1}(\psi(\qa))$ we have
$\psi(\qa_0') = \psi(\qa) = \psi(\qa_0'')$,
i.e.
$h_{\qa_0'}(\gamma_x) = h_\qa(\gamma_x) = h_{\qa_0''}(\gamma_x)$ for all $x$.
\item
Furthermore, we have
\bgl
          f(\varphi({\qa'}\circ\qg')) 
    & = & f(\varphi({\qa'_0}\circ\qz'\circ\qg')) \\
    & = & f(\varphi({\qa'_0})\circ z_m'\circ g_m') 
          \erl{$\varphi$ "equivariant"} \\
    & = & f(\varphi({\qa'_0}))\circ z_m'\circ g_m'
          \erl{$f$ equivariant} \\
    & = & \varphi({\qa})\circ z_m'\circ g_m'
          \erl{$\varphi({\qa'_0})\in S$} \\
    & = & \varphi({\qa}\circ \qz')\circ g_m'
          \erl{$\varphi$ "equivariant"} \\
    & = & \varphi({\qa})\circ g_m'
          \erl{$\qz'\in\bz(\qa)$}
\egl
and analogously $f(\varphi({\qa''}\circ\qg'')) = \varphi({\qa})\circ g_m''$.

Therefore, we have $\varphi(\qa)\circ g_m' = \varphi(\qa)\circ g_m''$, 
i.e. $g_m''\: (g'_m)^{-1}$ is an element of the stabilizer of $\varphi(\qa)$,
thus $g_m''\: (g'_m)^{-1}\in Z(\varphi(\qa)) = Z(\holgr_\qa)$.
\item
Since $\qa_0'\circ\qz'\circ\qg' = \qa''_0\circ\qz''\circ\qg''$, we have 
$\qa_0' = \qa''_0\circ\bigl(\qz''\:\qg''\:(\qg')^{-1}\:(\qz')^{-1}\bigr)$,
and so for all $x\in M$
\zgl{h_{\qa'_0}(\gamma_x) =
     \bigl(\qz''\:\qg''\:(\qg')^{-1}\:(\qz')^{-1}\bigr)_m^{-1} 
     \: h_{\qa''_0}(\gamma_x) \: 
     \bigl(\qz''\:\qg''\:(\qg')^{-1}\:(\qz')^{-1}\bigr)_x.}
Moreover, since $\bigl(\qg''\:(\qg')^{-1}\bigr)_m\in Z(\holgr_\qa)$, we have  
$\bigl(\qz''\:\qg''\:(\qg')^{-1}\:(\qz')^{-1}\bigr)_m\in Z(\holgr_\qa)$.
From $h_{\qa_0'}(\gamma_x) = h_\qa(\gamma_x) = h_{\qa_0''}(\gamma_x)$
for all $x$ now
$\qz''\:\qg''\:(\qg')^{-1}\:(\qz')^{-1}\in\bz(\qa)$ follows,
and thus
$\qg''\:(\qg')^{-1}\in\bz(\qa)$.
\item
By this we have $\qa\circ\qg' = \qa\circ\qg''$, i.e. $F$ is well-defined.
\eunum
\item
$F$ is equivariant.

\bunum
\item
Let $\qa'' = \qa'\circ\qg'\in\qS\circ\Gb$. Then
\bgl
        F(\qa''\circ\qg) 
  & = & F(\qa'\circ (\qg'\circ\qg)) \\
  & = & \qa\circ(\qg'\circ\qg)\\
  & = & (\qa\circ\qg')\circ\qg\\
  & = & F(\qa'\circ\qg')\circ \qg\\
  & = & F(\qa'') \circ\qg.
\egl  
\eunum
\item
$F$ is retracting.
\bunum
\item
Let $\qa' = \qa\circ\qg\in\qa\circ\Gb$. Then
$F(\qa') = F(\qa\circ\qg) = \qa\circ\qg = \qa'$.
\eunum
\item
$\qS\circ\Gb$ is an open neighbourhood of $\qa\circ\Gb$.
\bunum
\item
Obviously, $\qa\circ\Gb \teilmenge \qS\circ\Gb$.
\item
We have $\qS\circ\Gb = \varphi^{-1}(S\circ\LG)$.
\begin{SubSet}
Let $\qa'' = \qa' \circ\qg\in\qS_0\circ\Gb = \qS\circ\Gb$.

Then we have $\varphi(\qa'') = \varphi(\qa'\circ\qg) = \varphi(\qa') \circ g_m
\in S\circ\LG$ because $\varphi(\qS_0)\teilmenge S$. Thus, 
$\qa''\in\varphi^{-1} (S\circ\LG)$.
\end{SubSet}
\begin{SuperSet}
\bunum
\item
Let $\qa''\in\varphi^{-1}(S\circ\LG)$, i.e. $\varphi(\qa'') = \vec g''\circ
g$ with appropriate $\vec g''\in S$ and $g\in\LG$.
\item
Choose some $\qg$ with $g_m = g$. 

Then 
$\varphi(\qa''\circ\qg^{-1}) = \varphi(\qa'')\circ g_m^{-1} = \vec g''\in S$.

Now set $\qa''':=\qa''\circ\qg^{-1}$.
\item
Using
$g_x' := \bigl(h_{\qa'''}(\gamma_x)\bigr)^{-1} \: h_\qa(\gamma_x)$ and
$\qa':=\qa'''\circ\qg'$ we get
\bnum{2}
\item
$\varphi(\qa') = \varphi(\qa''') \in S$ because of $g_m'=e_\LG$ and
\item
$h_{\qa'}(\gamma_x) = h_{\qa'''}(\gamma_x) \: g_x' = h_\qa(\gamma_x)$
for all $x\in M$.
\enum
Thus, we have $\qa'\in\qS_0\teilmenge\qS$ and
$\qa'' = \qa'''\circ\qg = \qa'\circ((\qg')^{-1}\circ\qg) \in \qS\circ\Gb$.
\eunum
\end{SuperSet}
\item
Consequently, $\qS\circ\Gb = \varphi^{-1}(S\circ\LG)$ is as a preimage 
of an open set again open because of the continuity of $\varphi$.
\eunum
\item
$F$ is continuous.
\bunum
\item
We consider the following diagram
\znumgl{\begin{minipage}{0.77\textwidth}
\begin{center}
\vspace*{\CDgap}
\begin{minipage}{7.8cm}
\begin{diagram}[labelstyle=\scriptstyle,height=\CDhoehe,l>=3em]
\qS\circ\Gb            & \relax\rnach^{F} & \qa\circ\Gb            & & \\
\relax\dnach_{\varphi} &                  & \relax\dnach^{\varphi} & & \\
S\circ\LG              & \relax\rnach^{f} & \varphi(\qa)\circ\LG
                          & \relax\rnach^{\tau_\LG}_{\iso} & \nklza
\end{diagram}
\end{minipage}.
\end{center}
\end{minipage}
\label{cd1}} 
\begin{center}
\vspace*{\CDgap}
\begin{minipage}{8.4cm}
\begin{diagram}[labelstyle=\scriptstyle,height=\CDhoehe,l>=3em]
\qa'\circ\qg           & \relax\rauf^{F} & \qa\circ\qg            & & \\
\relax\dauf_{\varphi}  &                 & \relax\dauf^{\varphi}  & & \\
\varphi(\qa')\circ g_m & \relax\rauf^{f} & \varphi(\qa)\circ g_m
                          & \relax\rauf^{\tau_\LG} & [g_m]_{Z(\holgr_\qa)}
\end{diagram}
\end{minipage}
\end{center}
It is commutative due to 
$\varphi(\qS\circ\Gb) \teilmenge S\circ\LG$,
$\varphi(\qa\circ\Gb) \teilmenge \varphi(\qa)\circ\LG$ and
the definition of $F$.
$\tau_\LG$ is the canonical homeomorphism between the
orbit of $\varphi(\qa)$ and the quotient of the acting
group $\LG$ by the stabilizer of $\varphi(\qa)$.

Since $\varphi$, $f$ and $\tau_\LG$ are continuous, the map
\fktdefabgesetzt{F' := \tau_\LG\circ\varphi\circ F}{\qS\circ\Gb}{\nklza}
                         {\qa'\circ\qg}{[g_m]_{Z(\holgr_\qa)}}
is continuous.                         
\item
Now, we consider the map
\fktdefabgesetzt{F''}{(\qS\circ\Gb)\kreuz\LG}{\Gb.}
                     {(\qa'\circ\qg', g_m)}
                     {\bigl(h_{\gamma_x}(\qa)^{-1} \: g_m 
                        \: h_{\gamma_x}(\qa'\circ\qg')\bigr)_{x\in M}}
$F''$ is continuous because

$\begin{array}[t]{cccccc}%
  \pi_x\circ F'': & (\qS\circ\Gb)\kreuz\LG & \nach & \LG\kreuz\LG 
                  & \txtnach{\text{mult.}} & \LG\\    
                  & (\qa'', g_m)           & \auf  & (h_{\gamma_x}(\qa''), g_m) 
                  & \txtauf{\text{mult.}}  & h_{\gamma_x}(\qa)^{-1} \: g_m 
                                             \: h_{\gamma_x}(\qa'') 
\end{array}$

is obviously continuous for all $x\in M$.
\item
$F''$ induces a map $F'''$ via the following commutative diagram
\begin{center}
\vspace*{\CDgap}
\begin{minipage}{7.2cm}
\begin{diagram}[labelstyle=\scriptstyle,height=\CDhoehe,l>=3em]
(\qS\circ\Gb)\kreuz\LG    & \relax\rnach^{F''}  & \Gb                           \\
\relax\dnach^{\id\kreuz\pi_{Z(\holgr_\qa)}} 
                          &                     & \relax\dnach_{\pi_{\bz(\qa)}} \\
(\qS\circ\Gb)\kreuz\nklza & \relax\rnach^{F'''} & \nkla
\end{diagram}
\end{minipage},
\end{center}
i.e., $F'''(\qa'', [g_m]_{Z(\holgr_\qa)}) = 
        \bigl[\bigl(h_{\gamma_x}(\qa)^{-1} \: g_m 
                         \: h_{\gamma_x}(\qa'')\bigr)_{x\in M}
                            \bigr]_{\bz(\qa)}$.
\bunum
\item
$F'''$ is well-defined.

Let $g_{2,m} = z g_{1,m}$ with $z\in Z(\holgr_\qa)$.
Then 
\bgl
       F'''(\qa'', [g_{2,m}]_{Z(\holgr_\qa)})
 & = & \bigl[\bigl(h_{\gamma_x}(\qa)^{-1} \: g_{2,m} \:
        h_{\gamma_x}(\qa'')\bigr)_{x\in M}\bigr]_{\bz(\qa)} \\
 & = & \bigl[\bigl(h_{\gamma_x}(\qa)^{-1} \: z \: g_{1,m} \:
        h_{\gamma_x}(\qa'')\bigr)_{x\in M}\bigr]_{\bz(\qa)} \\
 & = & \bigl[\bigl(z_x \: h_{\gamma_x}(\qa)^{-1} \: g_{1,m} \:
        h_{\gamma_x}(\qa'')\bigr)_{x\in M}\bigr]_{\bz(\qa)} \\
 & = & F'''(\qa'', [g_{1,m}]_{Z(\holgr_\qa)}),
\egl
because
$(z_x)_{x\in M} := (h_{\gamma_x}(\qa)^{-1} \: z \: h_{\gamma_x}(\qa))_{x\in M}
                \in \bz(\qa)$ for $z\in Z(\holgr_\qa)$.
\item
$F'''$ is continuous, because $\id\kreuz\pi_{Z(\holgr_\qa)}$ is 
open and surjective and $\pi_{\bz(\qa)}$ and $F''$ are continuous.
\eunum
\item
For $\qa'\in\qS$ there is an $\qa_0'\in\qS_0$ and a $\qg'\in\bz(\qa)$
with $\qa' = \qa'_0\circ\qg'$. Thus, we have 
$h_{\gamma_x}(\qa_0') = h_{\gamma_x}(\qa)$ and
\bgl
       F'''(\qa'\circ\qg, [g_m])
 & = & \bigl[\bigl(h_{\gamma_x}(\qa)^{-1} \: g_m \:
        h_{\gamma_x}(\qa_0'\circ\qg'\circ\qg)\bigr)_{x\in M}\bigr]_{\bz(\qa)} \\
 & = & \bigl[\bigl(h_{\gamma_x}(\qa)^{-1} \: g_m \: g_m^{-1} (g'_m)^{-1}
        h_{\gamma_x}(\qa) g'_x g_x \bigr)_{x\in M}\bigr]_{\bz(\qa)} \\
 & = & \bigl[\bigl(h_{\gamma_x}(\qa)^{-1}
        h_{\gamma_x}(\qa\circ g') \: g_x \bigr)_{x\in M}\bigr]_{\bz(\qa)} \\
 & = & \bigl[(g_x)_{x\in M}\bigr]_{\bz(\qa)} \\
 & = & [\qg]_{\bz(\qa)}
\egl
where we used $\qg'\in\bz(\qa)$.
\item
Now, $F$ is the concatenation of the following continuous maps:

$\begin{array}[t]{cccccccc}%
   F: & \qS\circ\Gb               & \txtnach{\id\kreuz F'} 
      & (\qS\circ\Gb)\kreuz\nklza & \txtnach{F'''} 
      & \text{$\nkla$}            & \txtnach{\tau_{\Gb}}
      & \qa\circ\Gb,                                                   \\
      & \qa'\circ\qg              & \txtauf{\phantom{\id\kreuz F'}}
      & (\qa'\circ\qg, [g_m]_{Z(\holgr_\qa)}) & \txtauf{\phantom{F'''}}
      & [\qg]_{\bz(\qa)}          & \txtauf{\phantom{\tau_{\Gb}}}
      & \qa\circ\qg 
\end{array}$

where $\tau_\Gb$ is the canonical homeomorphism 
between the orbit $\qa\circ\Gb$ and the acting group $\Gb$ modulo the 
stabilizer $\bz(\qa)$ of $\qa$. 

Hence, $F$ is continuous.
\eunum
\item
We have $F^{-1}(\{\qa\}) = \qS$.
\bunum
\item
\begin{SubSet}
Let $\qa'\in F^{-1}(\{\qa\})$, i.e. $F(\qa') = \qa$.
\bunum
\item
By the commutativity of \eqref{cd1} we have 
$f(\varphi(\qa')) = \varphi(F(\qa')) = \varphi(\qa)$, hence
$\qa'\in\varphi^{-1}(f^{-1}(\varphi(\qa))) = \varphi^{-1}(S)$.
\item
Define $g_x := h_{\qa'}(\gamma_x)^{-1}\ h_\qa(\gamma_x)$ and
$\qa'':=\qa'\circ\qg$.
Then we have $\varphi(\qa'') = \varphi(\qa') \in S$, i.e. $\qa''\in\varphi^{-1}(S)$,
and $h_{\qa''}(\gamma_x) = h_\qa(\gamma_x)$ for all $x$, i.e.
$\qa''\in \psi^{-1}(\psi(\qa))$.
By this, $\qa''\in\qS_0$.
\item
Consequently, $F(\qa'') = \qa = F(\qa')$ and therefore also
$\qa\circ\qg = F(\qa') \circ \qg = F(\qa'\circ\qg) = F(\qa'') = \qa$, 
i.e. $\qg\in\bz(\qa)$.
\eunum
Thus, $\qa' = \qa''\circ\qg^{-1}\in\qS_0\circ\bz(\qa) = \qS$.
\end{SubSet}
\begin{SuperSet}
Let $\qa'\in\qS$. Then $F(\qa') = F(\qa'\circ 1) = \qa\circ 1 = \qa$,
i.e. $\qa'\in F^{-1}(\{\qa\})$.
\qed
\end{SuperSet}
\eunum
\enum
\epf

\section{Openness of the Strata}
\bprop
\label{Abtoffen}
$\Abgeq{t}$ is open for all $t\in\T$.
\eprop
\bcorr
$\Abeq t$ is open in $\Ableq t$ for all $t\in\T$.
\ecorr
\bpf
Since $\Abeq t = \Abgeq t \cap \Ableq t$, $\Abeq t$ is open
w.r.t. to the relative topology on $\Ableq t$.
\qed
\epf
\bcorr
\label{agbleqcompact}
$\Ableq t$ is compact for all $t\in\T$.
\ecorr
\bpf
$\Ab\setminus\Ableq t = \bigcup_{t'\in\T, t'\nichtleq t} \Abeq{t'}
                      = \bigcup_{t'\in\T, t'\nichtleq t} \Abgeq{t'}$
is open because $\Abgeq{t'}$ is open for all $t'\in\T$.
Thus, $\Ableq t$ is closed and therefore compact.
\qed
\epf
The proposition on the openness of the strata can be proven in two
ways: first as a simple corollary of the slice theorem on $\Ab$, 
but second directly using the reduction mapping. Thus, altogether
the second variant needs less effort.
\bpf[Proposition \ref{Abtoffen}]
We have to show that any $\qa\in\Abgeq t$ has a neighbourhood that 
again is contained in $\Abgeq t$. So, let $\qa\in\Abgeq t$.
\bunum
\item
Variant 1

Due to the slice theorem there is an open neighbourhood $U$ of $\qa\circ\Gb$,
and so of $\qa$, too, and an equivariant retraction $F:U\nach\qa\circ\Gb$.
Since every equivariant mapping reduces types, we have
$\typ(\qa')\geq\typ(\qa)=t$ for all $\qa'\in U$, thus $U\teilmenge\Abgeq t$.
\item
Variant 2

Choose again for $\qa$ an $\ga\teilmenge\hg$ with
\zgl{\typ(\qa) = [Z(\holgr_\qa)] = [Z(h_\qa(\ga))] \ident [Z(\varphi_\ga(\qa))]
               = \typ(\varphi_\ga(\qa)).}
Due to the slice theorem for general transformation groups
there is an open, invariant 
neighbourhood $U'$ of $\varphi_\ga(\qa)$ in $\LG^{\elanz\ga}$
and an equivariant retraction $f:U'\nach\varphi_\ga(\qa)\circ\LG$.
Since $\varphi_\ga(\qa)$ and $f$ are type-reducing,
we have 
\zgl{\typ(\qa') \geq \typ(\varphi_\ga(\qa'))
                \geq \typ\bigl(f(\varphi_\ga(\qa'))\bigr)
                 =   \typ(\varphi_\ga(\qa))
                 =   \typ(\qa)}
for all $\qa'\in U:=\varphi_\ga^{-1}(U')$,
i.e. $U\teilmenge\Abgeq t$. Obviously, $U$ contains $\qa$ and is 
open as a preimage of an open set.
\qed
\eunum
\epf

\section{Denseness of the Strata}
The next theorem we want to prove is that the set $\Abeq t$ is not only open,
but also dense in $\Ableq t$.
This assertion does -- in contrast to the slice theorem and the openness of the
strata -- not follow from the general theory of transformation groups.
We have to show this directly on the level of $\Ab$.

As we will see in a moment, the next proposition will be very helpful.
\bprop
\label{satz:ex_qa'_typ_t}
Let $\qa\in\Ab$ and $\GR_i$ be finitely many graphs.

Then there is for any $t\geq\typ(\qa)$ an $\qa'\in\Ab$
with $\typ(\qa') = t$ and $\pi_{\GR_i}(\qa) = \pi_{\GR_i}(\qa')$ for all $i$.
\eprop
Namely, we have
\bcorr
\label{folg:Ableq_dicht}
$\Abeq t$ is dense in $\Ableq t$ for all $t\in\T$.
\ecorr
\bpf
Let $\qa\in\Ableq t\teilmenge\Ab$.
We have to show that any neighbourhood $U$ of $\qa$ contains
an $\qa'$ having type $t$. It is sufficient to prove this assertion
for all graphs $\GR_i$ and all $U=\bigcap_i\pi_{\GR_i}^{-1}(W_i)$ with open
$W_i\teilmenge\LG^{\elanz\Edg(\GR_i)}$ and $\pi_{\GR_i}(\qa)\in W_i$ for all 
$i\in I$ with finite $I$, 
because any general open $U$ contains such a set.

Now let $\GR_i$ and $U$ be chosen as just described.
Due to Proposition \ref{satz:ex_qa'_typ_t} above there exists an
$\qa'\in\Ab$ with $\typ(\qa') = t \geq \typ(\qa)$ and
$\pi_{\GR_i}(\qa) = \pi_{\GR_i}(\qa')$ for all $i$, i.e.
with $\qa'\in\Abeq t$ and 
$\qa'\in\pi_{\GR_i}^{-1}\bigl(\pi_{\GR_i}(\{\qa\})\bigr)
     \teilmenge\pi_{\GR_i}^{-1}(W_i)$ for all $i$,
thus, $\qa'\in\bigcap_i \pi_{\GR_i}^{-1}(W_i)= U$.
\qed
\epf
Along with the proposition about the openness of the strata we get
\bcorr
\label{folg:abschl(Abeqt)}
For all $t\in\T$ the closure of $\Abeq t$ w.r.t. $\Ab$ is equal to $\Ableq t$.
\ecorr
\bpf
Denote the closure of $F$ w.r.t. $E$ by $\clos E(F)$.

Due to the denseness of $\Abeq t$ in $\Ableq t$ we have
$\clos{\Ableq t}(\Abeq t) = \Ableq t$.
Since the closure is compatible with the relative topology, we have
$\Ableq t = \clos{\Ableq t}(\Abeq t) = \Ableq t \cap \clos{\Ab}(\Abeq t)$,
i.e.
$\Ableq t \teilmenge \clos{\Ab}(\Abeq t)$.
But, due to Corollary \ref{agbleqcompact},
$\Ableq t\obermenge\Abeq t$ itself is
closed in $\Ab$. Hence, $\Ableq t \obermenge \clos{\Ab}(\Abeq t)$.
\qed
\epf
\subsection{How to Prove Proposition \ref{satz:ex_qa'_typ_t}?}
Which ideas will the proof of Proposition \ref{satz:ex_qa'_typ_t} be based on?
As in the last two sections 
we get help from the finiteness lemma for centralizers.
Namely, let $\ga\teilmenge\hg$ be chosen such that
$\typ(\qa) = [Z(\holgr_\qa)] = [Z(\varphi_\ga(\qa))]$. $t\geq\typ(\qa)$ 
is finitely generated as well.
Thus, we have to construct a connection whose type is determined by
$\varphi_\ga(\qa)$ and the generators of $t$.
For this we use the induction on the number of generators of $t$.
In conclusion, we have to construct inductively from $\qa$ new connections
$\qa_i$, such that $\qa_{i-1}$ coincides with $\qa_{i}$ at least along
the paths that pass $\ga$ or that lie in the graphs $\GR_i$.
But, at the same time, there has to exist a path $e$, such that $h_{\qa_{i}}(e)$
equals the $i$th generator of $t$.

Now, it should be obvious that we get help from the construction
method for new connections introduced in \cite{paper3}.
Before we do this we recall an important notation used there.
\bdf
Let $\gamma_1, \gamma_2\in\Pf$.

We say that $\gamma_1$ and $\gamma_2$ have the same initial segment 
(shortly: $\gamma_1 \BB \gamma_2$) iff there exist $0<\delta_1,\delta_2\leq 1$
such that
$\gamma_1\einschr{[0,\delta_1]}$ and $\gamma_2\einschr{[0,\delta_2]}$ 
coincide up to the parametrization.

We say analogously that the final segment of $\gamma_1$ coincides
with the initial segment of $\gamma_2$ 
(shortly: $\gamma_1 \EB \gamma_2$) iff there exist $0<\delta_1,\delta_2\leq 1$
such that
$\gamma_1^{-1}\einschr{[0,\delta_1]}$ and $\gamma_2\einschr{[0,\delta_2]}$ 
coincide up to the parametrization.

Iff the corresponding relations are not fulfilled, we write
$\gamma_1 \notBB \gamma_2$ and $\gamma_1 \notEB \gamma_2$, respectively.
\edf
Finally, we recall the decomposition lemma.
\blem
\label{zerleg_weg}
Let $x\in M$ be a point.
Any $\gamma\in\Pf$ can be written (up to parametrization)
as a product $\prod\gamma_i$ with $\gamma_i\in\Pf$,
such that
\bunum
\item
$\inter\gamma_i \cap \{x\} = \leeremenge$ or
\item
$\inter\gamma_i = \{x\}$.
\eunum
\elem

\subsection{Successive Magnifying of the Types}
In order to prove Proposition \ref{satz:ex_qa'_typ_t} we need the 
following lemma for magnifying the types.
Hereby, we will use explicitly the construction
of a new connection $\qa'$ from $\qa$ as given in \cite{paper3}.
\blem
\label{folg:step1(typ)}
Let $\GR_i$ be finitely many graphs,
$\qa\in\Ab$ and $\ga\teilmenge\hg$ be a finite set of paths with
$Z(\holgr_\qa) = Z(h_\qa(\ga))$. Furthermore, let $g\in\LG$ be
arbitrary.

Then there is an $\qa'\in\Ab$, such that:
\bunum
\item
$h_{\qa'}(\ga) = h_\qa(\ga)$,
\item
$\pi_{\GR_i}(\qa') = \pi_{\GR_i}(\qa)$ for all $i$,
\item
$h_{\qa'}(e) = g$ for an $e\in\hg$ and
\item
$Z(\holgr_{\qa'}) = Z(\{g\}\cup h_\qa(\ga))$.
\eunum
\elem
\bpf
\bnum{2}
\item
Let $m'\in M$ be some point that is neither contained in the images of $\GR_i$ 
nor in that of $\ga$, and join $m$ with $m'$ by some path $\gamma$. 
Now let $e'$ be some closed path in $M$ with base point $m'$
and without self-intersections,
such that 
\znumgl{\label{bed:loop_e}
        \im e' \cap 
        \bigl(\inter\gamma\cup\im(\ga)\cup\bigcup\im(\GR_i)\bigr)\bigr) 
        = \leeremenge.}
Obviously, there exists such an $e'$ because $M$ is supposed to be at least
two-dimensional. Set $e := \gamma \: e' \: \gamma^{-1}\in\hg$ and
$g':=h_\qa(\gamma)^{-1} g h_\qa(\gamma)$.

Finally, define a connection $\qa'$ for $\qa$, $e'$ and $g'$
as follows:
\item
Construction of $\qa'$

\bunum
\item
Let $\delta\in\Pf$ be for the moment a "genuine" path (i.e., not
an equivalence class) that does not contain the
initial point $\bweg'(0)\ident m'$ of $\bweg'$ 
as an inner point. Explicitly we have
$\inter\delta \cap \{\bweg'(0)\} = \leeremenge$.
Define

$h_{\qa'}(\delta) :=
\begin{cases}
  g' \: h_\qa(\bweg')^{-1} \: 
        h_\qa(\delta) \: h_\qa(\bweg') \: {g'}^{-1}, 
                 & \text{ for $\delta \BB \bweg'$ and $\delta \EB \bweg'$} \\
  g' \: h_\qa(\bweg')^{-1} \: 
        h_\qa(\delta)\phantom{ \: h_\qa(\bweg') \: {g'}^{-1}}, 
                 & \text{ for $\delta \BB \bweg'$ and $\delta \notEB \bweg'$} \\
  \phantom{g' \: h_\qa(\bweg')^{-1} \: 
       }h_\qa(\delta) \: h_\qa(\bweg') \: {g'}^{-1}, 
                 & \text{ for $\delta \notBB \bweg'$ and $\delta \EB \bweg'$} \\
  \phantom{g' \: h_\qa(\bweg')^{-1} \: }h_\qa(\delta)\phantom{ \: h_\qa(\bweg') 
    \: {g'}^{-1}},  & \text{ else}        
\end{cases}$ .
\item
For every trivial path $\delta$ set $h_{\qa'}(\delta) = e_\LG$.
\item
Now, let $\delta\in\Pf$ be an arbitrary path. Decompose $\delta$ into a 
finite product $\prod\delta_i$ due to Lemma \ref{zerleg_weg} such that no
$\delta_i$ contains the point $\bweg'(0)$ in the interior supposed 
$\delta_i$ is not trivial.
Here, set $h_{\qa'}(\delta) := \prod h_{\qa'}(\delta_i)$.
\eunum
We know from \cite{paper3} that $\qa'$ is indeed a connection.
\item
The assertion $\pi_{\GR_i}(\qa')=\pi_{\GR_i}(\qa)$ for all $i$
is an immediate consequence of
the construction because
$\im(\GR_i) \cap \inter e' = \leeremenge$. As well, we get
$h_{\qa'}(\ga) = h_{\qa}(\ga)$.
\item
Moreover, from \eqref{bed:loop_e}, 
the fact that $e'$ has no self-intersections and the 
definition of $\qa'$ we get $h_{\qa'}(\gamma) = h_\qa(\gamma)$ and so
\zgl{h_{\qa'}(e) = h_{\qa'}(\gamma) \: h_{\qa'}(e') \: h_{\qa'}(\gamma^{-1})
             = h_{\qa}(\gamma) \: g' \: h_{\qa}(\gamma)^{-1} = g.}
\item
We have $Z(\holgr_{\qa'}) = Z(\{g\}\cup\holgr_\qa)$.
\begin{SubSet}
Let $f\in Z(\holgr_{\qa'})$, i.e. $f\:h_{\qa'}(\alpha) = h_{\qa'}(\alpha) \: f$
for all $\alpha\in\hg$.
\bunum
\item
From $h_{\qa'}(e) = g$ follows $fg=gf$, i.e.
$f\in Z(\{g\})$.
\item
From $\im e'\cap\im(\ga)=\leeremenge$
follows $h_\qa(\alpha_i) = h_{\qa'}(\alpha_i)$, i.e.
$f\in Z(h_\qa(\alpha_i))$ for all $i$.
\eunum
Thus, $f\in Z(\{g\}) \cap Z(h_\qa(\ga)) = Z(\{g\}\cup\holgr_\qa)$.
\end{SubSet}
\begin{SuperSet}
Let $f\in Z(\{g\}\cup \holgr_\qa)$.
\bunum
\item
Let $\alpha'$ be a path from $m'$ to $m'$, such that 
$\inter\alpha'\cap \{m'\}=\leeremenge$ or
$\inter\alpha' = \{m'\}$. 
Set $\alpha:=\gamma\:\alpha'\:\gamma^{-1}$.
Then by construction we have 
\bgl
h_{\qa'}(\alpha) & = & h_{\qa'}(\gamma) \: h_{\qa'}(\alpha') \:
                       h_{\qa'}(\gamma)^{-1} \\
                 & = & h_\qa(\gamma) \: h_{\qa'}(\alpha') \: h_\qa(\gamma)^{-1}.
\egl               
There are four cases:
\bunum
\item
$\alpha'\notBB e'$ and $\alpha'\notEB e'$: 
\bgl
h_{\qa'}(\alpha) & = & h_\qa(\gamma) \: h_\qa(\alpha') \: h_\qa(\gamma)^{-1}
                 \breitrel= h_\qa(\gamma\:\alpha'\:\gamma^{-1}) \\
                 & = & h_\qa(\alpha).
\egl                 
\item
$\alpha'\BB e'$ and $\alpha'\notEB e'$: 
\bgl
h_{\qa'}(\alpha) & = & h_\qa(\gamma) \:
                       g' \:h_\qa(e')^{-1} \:h_\qa(\alpha') \:
                       h_\qa(\gamma)^{-1}  \\
                 & = & g \:h_\qa(\gamma) \:
                       h_\qa(e')^{-1} \:h_\qa(\alpha') \:
                       h_\qa(\gamma)^{-1} \\
                 & = & g \:h_\qa(\gamma {e'}^{-1}\alpha' \gamma^{-1}).
\egl
\item
$\alpha'\notBB e'$ and $\alpha'\EB e'$: 
\bgl 
h_{\qa'}(\alpha) & = & h_\qa(\gamma) \:
                       h_\qa(\alpha') \: h_\qa(e') \: (g')^{-1}
                       h_\qa(\gamma)^{-1} \\
                 & = & h_\qa(\gamma) \:
                       h_\qa(\alpha')  \: h_\qa(e') 
                       h_\qa(\gamma)^{-1} \: g^{-1}\\
                 & = & h_\qa(\gamma \alpha' e' \gamma^{-1}) \: g^{-1}.
\egl                  
\item
$\alpha'\BB e'$ and $\alpha'\EB e'$: 
\bgl 
h_{\qa'}(\alpha) & = & h_\qa(\gamma) \:
                       g' \: h_\qa(e')^{-1} \: 
                       h_\qa(\alpha') \: h_\qa(e') \: (g')^{-1} \:
                       h_\qa(\gamma)^{-1} \\
                 & = & g \: h_\qa(\gamma) \: h_\qa(e')^{-1} \:
                       h_\qa(\alpha') \: h_\qa(e') \:
                       h_\qa(\gamma)^{-1} \: g^{-1}\\
                 & = & g \: h_\qa(\gamma {e'}^{-1} \alpha' e' \gamma^{-1}) \:
                       g^{-1}.
\egl
\eunum
Thus, in each case we get $f\in Z(\{h_{\qa'}(\alpha)\})$.
\item
Now, let $\alpha\in\hg$ be arbitrary and $\alpha':=\gamma^{-1}\alpha\gamma.$

By the Decomposition Lemma \ref{zerleg_weg}
there is a decomposition $\alpha' = \prod \alpha'_i$ with
$\inter\alpha_i'\cap \{m'\}=\leeremenge$ or
$\inter\alpha'_i = \{m'\}$ for all $i$.
Thus, $\alpha = \gamma \bigl(\prod \alpha'_i\bigr) \gamma^{-1}
                 = \prod\bigl(\gamma \alpha'_i \gamma^{-1}\bigr)$.
Using the result just proven  we get
$f\in Z\bigl(\bigl\{h_{\qa'}
         \bigl(\prod \bigl(\gamma \alpha'_i \gamma^{-1}\bigr)\bigr)\bigr\}\bigr)
    = Z(\{h_{\qa'}(\alpha)\})$.
\eunum
Thus, $f\in Z(\holgr_{\qa'})$.
\end{SuperSet}
Due to the definition of $\ga$ we have 
$Z(\holgr_{\qa'}) = Z(\{g\}\cup h_\qa(\ga))$.
\qed
\enum
\epf

\subsection{Construction of Arbitrary Types}
Finally, we can now prove the desired proposition.
\bpf[Proposition \ref{satz:ex_qa'_typ_t}]
\bunum
\item
Let $t\in\T$ and $t\geq\typ(\qa)$. 
Then there exist a Howe subgroup $V'\teilmenge\LG$
with $t = [V']$ and a $g\in\LG$, such that
$Z(\holgr_\qa) \obermenge g^{-1} V' g =: V$. Since $V$ is a 
Howe subgroup, we have $Z(Z(V)) = V$ and so by Lemma \ref{exkelemzentr}
there exist certain $u_0,\ldots,u_k\in Z(V)\teilmenge\LG$, such that
$V = Z(Z(V)) = Z(\{u_0,\ldots,u_k\})$.
\item
Now let $Z(\holgr_\qa) = Z(h_\qa(\ga))$ with an appropriate $\ga\teilmenge\hg$
as in Corollary \ref{endlbestzentr}. Because of
$V\teilmenge Z(\holgr_\qa)$ we have
$V = V \cap Z(\holgr_\qa) = Z(\{u_0,\ldots,u_k\})\cap Z(h_\qa(\ga))
   = Z(\{u_0,\ldots,u_k\} \cup h_\qa(\ga))$.
\item
We now use inductively Lemma \ref{folg:step1(typ)}.
Let $\qa_0:=\qa$ and $\ga_0:=\ga$.
Construct for all $j=0,\ldots, k$ 
a connection $\qa_{j+1}$ and an $e_j\in\hg$
from $\qa_j$ and $\ga_j$ by that lemma, such that
$\pi_{\GR_i}(\qa_{j+1}) = \pi_{\GR_i}(\qa_j)$ for all $i$, 
$h_{\qa_{j+1}}(\ga_j) = h_{\qa_j}(\ga_j)$,
$h_{\qa_{j+1}}(e_j) = u_j$ and
$Z(\holgr_{\qa_{j+1}}) = Z(\{u_j\}\cup h_{\qa_j}(\ga_j))$.

Setting $\ga_{j+1} := \ga_{j}\cup\{e_j\}$
we get 
$Z(\holgr_{\qa_{j+1}}) = Z(\{u_j\}\cup h_{\qa_j}(\ga_j)) 
                      = Z(h_{\qa_{j+1}}(\ga_{j+1}))$.
Finally, we define $\qa':=\qa_{k+1}$.

Now, we get $\pi_{\GR_i}(\qa') = \pi_{\GR_i}(\qa)$ for all $i$, 
$h_{\qa'}(\ga) = h_\qa(\ga)$ and
$h_{\qa'}(e_j) = u_j$. Thus,
\bgl
Z(\holgr_{\qa'}) & = & Z(h_{\qa'}(\ga_{k+1})) \\
                 & = & Z(h_{\qa'}(\{e_0,\ldots,e_k\}\cup h_{\qa'}(\ga))) \\
                 & = & Z(\{u_0,\ldots,u_k\}\cup h_\qa(\ga)) \\
                 & = & V,
\egl
i.e., $\typ(\qa') = [V] = t$.
\qed
\eunum
\epf
The proposition just proven has a further immediate consequence.
\bcorr
\label{folg:stratum_nie_leer}
$\Abeq t$ is non-empty for all $t\in\T$.
\ecorr
\bpf
Let $\qa$ be the trivial connection, i.e. $h_\qa(\alpha) = e_\LG$ for all
$\alpha\in\Pf$. The type of $\qa$ is $[\LG]$, thus minimal,
i.e. we have $t\geq\typ(\qa)$ for all $t\in\T$. 
By means of Proposition \ref{satz:ex_qa'_typ_t}
there is an $\qa'\in\Ab$ with $\typ(\qa')=t$.
\qed
\epf
This corollary solves the problem which gauge orbit types exist for 
generalized connections.
\bthm
The set of all gauge orbit types on $\Ab$ is the set
of all conjugacy classes of Howe subgroups of $\LG$.
\ethm
Furthermore we have
\bcorr
Let $\GR$ be some graph. Then $\pi_\GR(\Abeq{t_{\max}}) = \pi_\GR(\Ab)$.
In other words: $\pi_\GR$ is surjective even on the generic connections.
\ecorr
\bpf
$\pi_\GR$ is surjective on $\Ab$ as proven in \cite{paper3}.
By Proposition \ref{satz:ex_qa'_typ_t} there is now an $\qa'$ with
$\typ(\qa') = t_{\max}$ and $\pi_\GR(\qa') = \pi_\GR(\qa)$.
\qed
\epf

\section{Stratification of $\Ab$}
\label{abschn:stratifiz}
First we recall the general definition of a stratification \cite{f11}.
\bdf
A countable family $\stratallg$
of non-empty subsets of a topological
space $X$ is called \df{stratification} of $X$ iff 
$\stratallg$ is a covering for $X$ and for all $U,V\in\stratallg$
we have
\bunum
\item
$U\cap V\neq\leeremenge$ $\impliz$ $U=V$,
\item
$\quer U\cap V\neq\leeremenge$ $\impliz$ $\quer U\obermenge V$ and
\item
$\quer U\cap V\neq\leeremenge$ $\impliz$ $\quer V\cap(U\cup V) = V$.
\eunum
The elements of such a stratification $\stratallg$ are called
\df{strata}.

A stratification is called \df{topologically regular} iff
for all $U,V\in\stratallg$
\zgl{\text{$U\neq V$ and $\quer U\cap V\neq\leeremenge$ $\impliz$ 
$\quer V\cap U=\leeremenge$.}}
\edf
\bthm
$\strat:=\{\Abeq t\mid t\in\T\}$ is a
topologically regular stratification of $\Ab$.

Analogously, $\{(\AbGb)_{= t}\mid t\in\T\}$ is a 
topologically regular stratification of $\AbGb$.
\ethm
\bpf
\bunum
\item
Obviously, $\strat$ is a covering of $\Ab$.
\item
For a compact Lie group the set of all types, i.e. all conjugacy classes 
of Howe subgroups of $\LG$, is at most countable (cf. \cite{f11}).
\item
Moreover, from $\Abeq{t_1} \cap \Abeq{t_2} \neq \leeremenge$ immediately follows
$\Abeq{t_1} = \Abeq{t_2}$.
\item
Due to Corollary \ref{folg:abschl(Abeqt)} we have\footnote{$\clos{}(U)$ denotes
again the closure of $U$, here w.r.t. $\Ab$.}
$\clos{}(\Abeq{t_1}) = \Ableq{t_1}$, i.e.
from $\clos{}(\Abeq{t_1}) \cap \Abeq{t_2}\neq\leeremenge$ follows $t_2\leq t_1$
and thus $\clos{}(\Abeq{t_1})\obermenge\Abeq{t_2}$.
\item
Analogously we get $\clos{}(\Abeq{t_2})\cap(\Abeq{t_1}\cup\Abeq{t_2}) 
= \Ableq{t_2}\cap(\Abeq{t_1}\cup\Abeq{t_2}) = \Abeq{t_2}$.
\item
As well, from $\clos{}(\Abeq{t_1})\cap\Abeq{t_2}\neq\leeremenge$
and $\Abeq{t_1}\neq\Abeq{t_2}$ follows $t_1>t_2$, i.e. 
$\clos{}(\Abeq{t_2})\cap\Abeq{t_1}=\leeremenge$.
\eunum
Consequently, $\strat$ is a topologically regular stratification of $\Ab$.
\qed
\epf
For a regular stratification it would be required that each
stratum carries the structure of a manifold that is compatible with 
the topology of the total space. In contrast to the case of the 
classical gauge orbit space \cite{f11}, this is not fulfilled for generalized
connections.
\section{Non-complete Connections}
\label{section:noncomplconn}
We shall round off that paper with the proof that the set of the
so-called non-complete connections is contained in a set of measure zero.
This section actually stands a little bit separated from the context because
it is the only section that is not only algebraic and topological, but
also measure theoretical. 
\bdf
Let $\qa\in\Ab$ be a connection.
\bnum{3}
\item
$\qa$ is called \df{complete} $\aequ$ $\holgr_\qa = \LG$.
\item
$\qa$ is called \df{almost complete} $\aequ$ $\quer{\holgr_\qa} = \LG$.
\item
$\qa$ is called \df{non-complete} $\aequ$ $\quer{\holgr_\qa} \neq \LG$.
\enum
\edf
Obviously, we have 
\blem
If $\qa\in\Ab$ is complete (almost complete, non-complete), so
$\qa\circ\qg$ is complete (almost complete, non-complete) for all
$\qg\in\Gb$.
\elem
Thus, the total information about the completeness of a connection
is already contained in its gauge orbit.
Now, to the main assertion of this section.
\bprop
\label{nichtgenistnull}
Let $N := \{\qa\in\Ab\mid \qa\text{ non-complete}\}.$
Then $N$ is contained in a set of $\mu_0$-measure zero whereas
$\mu_0$ is the induced Haar measure on $\Ab$. \cite{a48,d3,paper3}
\eprop
Since $N$ is gauge invariant, we have
\bcorr
Let $[N] := \{[\qa]\in\AbGb\mid \qa\text{ non-complete}\}.$ 
Then $[N]$ is contained in a set of $\mu_0$-measure zero.
\ecorr
For the proof of the proposition we still need the following
\blem
\label{lemUinGnull}
Let $U\teilmenge\LG$ be measurable with $\mu_\Haar (U)>0$ and
$N_U := \{\qa\in\Ab\mid\holgr_\qa\teilmenge \LG\setminus U\}$.

Then $N_U$ is contained in a set of $\mu_0$-measure zero.
\elem
\bpf
\bunum
\item
Let $k\in\N$ and $\GR_k$ be some connected graph 
with one vertex $m$ and $k$ edges 
$\alpha_1,\ldots,\alpha_k\in\hg$.\footnote{Such a graph does
indeed exist for $\dim M\geq 2$. For instance, take 
$k$ circles $K_i$ with centers in $(\inv i,0,\ldots)$ and
radii $\inv i$. By means of an appropriate chart mapping around 
$m$ these circles define a graph with the desired properties.}
Furthermore, let 
\fktdef{\pi_k}{\Ab}{\LG^k.}{\qa}{(h_\qa(\alpha_1),\ldots,h_\qa(\alpha_k))}
\item
Denote now by $N_{k,U} := \pi_k^{-1} ((\LG\setminus U)^k)$ the set of all
connections whose holonomies on $\GR_k$ are not contained in $U$.
Per constructionem we have $N_U\teilmenge N_{k,U}$.
\item
Since the characteristic function $\chi_{N_{k,U}}$ for $N_{k,U}$ 
is obviously a cylindrical function, we get
\bgl[2ex]
\mu_0(N_{k,U}) & = & \int_\Ab \chi_{N_{k,U}} \: d\mu_0
               \breitrel= \int_\Ab \pi_k^\ast(\chi_{(\LG\setminus U)^k}) \: 
                            d\mu_0\s
                    & = & \int_{\LG^k} \chi_{(\LG\setminus U)^k} \: 
                            d\mu_\Haar^k
               \breitrel= [\mu_\Haar(\LG\setminus U)]^k.
\egl
\item
From $N_U\teilmenge N_{k,U}$ for all $k$ follows
$N_U\teilmenge \bigcap_k N_{k,U}$. But,  
$\mu_0(\bigcap_k N_{k,U}) \leq \mu_0(N_{k,U}) = 
 \mu_\Haar(\LG\setminus U)^k$ for all $k$, i.e.
$\mu_0(\bigcap_k N_{k,U}) = 0$, because 
$\mu_\Haar(\LG\setminus U) = 1 - \mu_\Haar(U)<1$.
\qed
\eunum
\epf
\bpf[Proposition \ref{nichtgenistnull}]
\bunum
\item
Let $(\epsilon_k)_{k\in\N}$ be some null sequence.
Furthermore, let $\{U_{k,i}\}_i$ be for each $k$ a finite covering of $\LG$
by open sets $U_{k,i}$ whose respective diameters are smaller than $\epsilon_k$.
Now define $N' := \bigcup_k \bigl(\bigcup_i N_{U_{k,i}} \bigr)$.
\item
Since $U_{k,i}$ is open and $\LG$ is 
compact, $U_{k,i}$ is measureable with $\mu_\Haar(U_{k,i})>0$.
Due to Lemma \ref{lemUinGnull} we have $N_{U_{k,i}}\teilmenge N^\ast_{U_{k,i}}$
with $\mu_0(N^\ast_{U_{k,i}}) = 0$ for all $k,i$; thus
$N'\teilmenge N^\ast:=\bigcup_k \bigl(\bigcup_i N^\ast_{U_{k,i}} \bigr)$
with $\mu_0(N^\ast) = 0$.
\item
We are left to show $N\teilmenge N'$.

Let $\qa\in N$. Then there is an open $U\teilmenge\LG$
with $\holgr_\qa\teilmenge \LG\setminus U$.

Now let $m\in U$. Then 
$\epsilon := \dist(m,\del U) > 0$. Choose $k$ such that $\epsilon_k<\epsilon$. 
Then choose a $U_{k,i}$ with $m\in U_{k,i}$. We get for all $x\in U_{k,i}$:
$d(x,m)\leq\diam U_{k,i}<\epsilon_k<\epsilon,$ i.e. $x\in U$.
Consequently, $U_{k,i}\teilmenge U$ and thus 
$\holgr_\qa\teilmenge \LG\setminus U_{k,i}$, i.e. $\qa\in N'$.
\qed
\eunum
\epf
\bcorr
The set of all generic connections (i.e. connections of maximal type)
has $\mu_0$-measure 1.
\ecorr
\bpf
Every almost complete connection $\qa$ has 
type $[Z(\holgr_\qa)] = [Z(\LG)] = t_{\max}$.
(Observe that the centralizer of a set $U\teilmenge\LG$ equals that of 
the closure
$\quer U$.) Since $\Abeq{t_{\max}}$ is open due to Proposition \ref{Abtoffen},
thus measurable, Proposition \ref{nichtgenistnull} yields the assertion.
\qed
\epf
The last assertion is very important: It justifies the definition
of the natural induced Haar measure on
$\AbGb$ (cf. \cite{a48,paper3}). Actually, 
there were (at least) two different possibilities for this.
Namely, let $X$ be some general topological space equipped with a
measure $\mu$ and let $G$ be some topological group acting on $X$. 
The problem now is to find a natural measure $\mu_G$
on the orbit space $X/G$. On the one hand, one could simply 
define $\mu_G(U) := \mu(\pi^{-1}(U))$ for all measurable $U\teilmenge X/G$. 
($\pi:X\nach X/G$ is the canonical projection.) But, on the other hand, one 
also could stratify the orbit space. For instance, in the easiest case
we could have $X = X/G \kreuz G$. 
In general, one gets (roughly speaking) 
$X = \bigcup \bigl(V/G \kreuz \rnkl{G_V}{G}\bigr)$
whereas $\bigcup V$ is an appropriate disjoint decomposition of $X$ and
$G_V$ characterizes the type of the orbits on $V$.
Now one naively defines
$\mu_G(U) := \sum_V \frac{\mu(\pi^{-1}(U)\cap V)}{\mu_{G,V}(G/G_V)}
          := \sum_V \mu\bigl(\pi^{-1}(U)\cap V\bigr) \mu_V(G_V)$,
where $\mu_V$ measures the "size" of the stabilizer $G_V$ in
$G$. This second variant is nothing but the transformation
of the measures using the Faddeev-Popov determinant (i.e. the Jacobi determinant)
$\frac{d\mu}{d\mu_G}$. 
In contrast to the first method, here the orbit space and not the total space
is regarded to be primary. 
For a uniform distribution of the measure over all points of the total space 
the image measure on the orbit space needs 
no longer be uniformly distributed; the orbits are weighted by size.
But, for the second method the uniformity is maintained. In other words,
the gauge freedom does not play any r\^ole when the Faddeev-Popov method
is used.

Nevertheless, we see in our concrete case of $\pi_\AbGb:\Ab\nach\AbGb$ 
that both methods are equivalent because the  
Faddeev-Popov determinant is equal to $1$ (at least outside a set
of $\mu_0$-measure zero). This follows immediately from the slice theorem
and the corollary above that the generic connections have total 
measure $1$. 
\section{Summary and Discussion}
In the present paper and its predecessor \cite{paper2} we gained
a lot of information about the structure of the generalized
gauge orbit space within the Ashtekar framework.
The most important tool was the theory of compact 
transformation groups on topological spaces. This enabled 
us to investigate the action of the group of generalized
gauge transforms on the space of generalized
connections. Our considerations were guided by the results of Kondracki and 
Rogulski \cite{f11} about the structure of the classical
gauge orbit space for Sobolev connections. The methods used there are
however fundamentally different from ours. Within the Ashtekar approach
most of the proofs are purely algebraic or topological;
in the classical case the methods are especially based on the 
theory of fiber bundles, i.e. analysis and differential geometry.

In a preceding paper \cite{paper2} we proved that the
$\Gb$-stabilizer $\bz(\qa)$ of a connection $\qa$ is isomorphic
to the $\LG$-centralizer $Z(\holgr_\qa)$ of the holonomy group of $\qa$.
Furthermore, two connections have conjugate $\Gb$-stabilizers if and only 
if their holonomy centralizers are conjugate. Thus, the type of
a generalized connection can be defined equivalently both by the
$\Gb$-conjugacy class of $\bz(\qa)$ (as known from the general theory
of transformation groups) and by the $\LG$-conjugacy class of 
$Z(\holgr_\qa)$. This is a significant difference to the classical 
case.

The reduction of our problem from structures in $\Gb$ to those in $\LG$
was the crucial idea in the present paper. Since stabilizers
in compact groups are even generated by a finite number of elements,
we could model the gauge orbit type $[Z(\holgr_\qa)]$
on a finite-dimensional space. Using an appropriate mapping
we lifted the corresponding slice theorem to a slice theorem
on $\Ab$. This is the main result of our paper. Collecting connections
of one and the same type we got the so-called strata whose openness
was an immediate consequence of the slice theorem. In the next step 
we showed that the natural ordering on the set of the types
encodes the topological properties
of the strata. More precisely, we proved that the closure of a stratum 
contains (besides the stratum itself) exactly the union of all strata
having a smaller type. This implied that this decomposition
of $\Ab$ is a topologically regular stratification.

All these results hold in the classical case as well. This
is very remarkable because our proofs used partially completely 
different ideas. However, two results of this paper 
go beyond the classical theorems. First, we were able to determine the
full set of all gauge orbit types occurring in $\Ab$. This
set is known for Sobolev connections -- to the best of our knowlegde -- 
only for certain bundles. Recently, Rudolph, Schmidt and Volobuev
solved this problem completely for $SU(n)$-bundels $P$ over two-, three-
and four-dimensional manifolds \cite{f14}.
The main problem in the Sobolev case is the non-triviality of the bundle $P$.
This can exclude orbit types that occur in the trivial bundle $M\kreuz SU(n)$.
But, this problem is irrelevant for the Ashtekar framework: {\em Every}
regular connection in {\em every} $\LG$-bundle over $M$ is contained in $\Ab$
\cite{a48}. This means, in a certain sense, we only have to deal with
trivial bundles.
Second, in the Ashtekar framework there is
a well-defined
natural measure on $\Ab$. 
Using this we could show that the generic stratum
has the total measure one; this is not true in the classical case.
The proposition above implies now that the Faddeev-Popov determinant
for the transformation from $\Ab$ to $\AbGb$ is equal to $1$.
This, on the other hand, justifies the definition of the induced Haar 
measure on $\AbGb$ by projecting the corresponding measure
for $\Ab$ which has been discussed in detail 
in section \ref{section:noncomplconn}.

Hence, we were able to "transfer" the classical theory of strata in a certain sense
(almost) completely to the Ashtekar program.
We emphasize that all assertions
are valid for {\em each} compact structure group -- {\em both}
in the analytical {\em and} in the $C^r$-smooth case.

\leerezeile

What could be next steps in this area? An important -- and in this 
paper completely ignored -- item is the physical interpretation of
the gained knowledge. So we will conclude our paper with a few
ideas that could link mathematics and physics:
\bunum
\item
{\em Topology}

What is the topological structure of the strata? Are they connected
or is $\Ab$ connected itself (at least for connected $\LG$)?
Is $\Abeq t$ globally trivial over $(\AbGb)_{=t}$, at least for the generic 
stratum with $t = t_{\max}$? What sections do exist in these bundles,
i.e. what gauge fixings do exist in $\Ab$?

These problems are closely related to the so-called
Gribov problem, the non-existence of global gauge fixings
for classical connections in principal fiber bundles
with compact, non-commutative structure group (see, e.g., \cite{f6}).
From this lots of difficulties result for the quantization
of such a Yang-Mills theory that are not circumvented up to now.
\item
{\em Algebraic topology}
\keinseitenumbr

Is there a meaningful, i.e. especially non-trivial cohomology theory
on $\Ab$?\footnote{First abstract attempts can be found, e.g., in 
\cite{a30,a28}.}
Is it possible to construct this way characteristic classes or
even topological invariants?
\item
{\em Measure theory}

How are arbitrary measures distributed over single strata? In other 
words: What properties do measures have that are defined by the
choice of a measure on each single stratum?

This is extremely interesting, in particular, 
from the physical point of view because the choice of a 
$\mu_0$-absolutely continuous measure $\mu$ on $\Ab$
corresponds to the choice of an action functional $S$ on $\Ab$ 
by $\int_\Ab f\: d\mu = \int_\Ab f\: e^{-S} \: d\mu_0$.
According to Lebesgue's decomposition theorem all
measures whose support is not fully contained in the generic stratum have
singular parts.
\eunum
Finally, we have to stress that the present paper only investigates
the case of pure gauge theories. Of course, this is physically not
satisfying. Therefore the next goal should be the inclusion of matter fields. 
A first step has already been done by Thiemann \cite{e35} whereas
the aspects considered in the present paper did not play any r\^ole in 
Thiemann's paper.

\section*{Acknowledgements}
I am very grateful to Gerd Rudolph and Eberhard Zeidler for their 
great support while I wrote my diploma thesis and the present paper.
Additionally, I thank Gerd Rudolph for reading the drafts.
Moreover, I am grateful to Domenico Giulini and Matthias Schmidt 
for convincing me to hope for the existence of a slice theorem on
$\Ab$. Finally, I thank the 
Max-Planck-Institut f\"ur Mathematik in den Naturwissenschaften
for its generous promotion. 

\addcontentsline{toc}{section}{Literaturverzeichnis}
\bibliographystyle{plain}

\end{document}